\documentclass[10pt,journal,compsoc]{IEEEtran}
\pdfoutput=1
\usepackage{graphicx,multicol} 
\usepackage{epsfig} 
\usepackage{amsmath,amsfonts,amsthm,bm} 
\usepackage{amssymb}  
\usepackage{float}
\usepackage{caption}
\usepackage{cases,setspace,adjustbox}
\usepackage{subfig}
\newlength{\twosubht}
\newsavebox{\twosubbox}
\usepackage{cite}
\usepackage{algorithm,algorithmicx,algpseudocode,enumitem}

\usepackage{etoolbox}

\makeatletter
\newcommand*{\algrule}[1][\algorithmicindent]{%
  \makebox[#1][l]{%
    \hspace*{.2em}
    \vrule height .75\baselineskip depth .25\baselineskip
  }
}

\newcount\ALG@printindent@tempcnta
\def\ALG@printindent{%
    \ifnum \theALG@nested>0
    \ifx\ALG@text\ALG@x@notext
    \else
    \unskip
    \ALG@printindent@tempcnta=1
    \loop
    \algrule[\csname ALG@ind@\the\ALG@printindent@tempcnta\endcsname]%
    \advance \ALG@printindent@tempcnta 1
    \ifnum \ALG@printindent@tempcnta<\numexpr\theALG@nested+1\relax
    \repeat
    \fi
    \fi
}
\patchcmd{\ALG@doentity}{\noindent\hskip\ALG@tlm}{\ALG@printindent}{}{\errmessage{failed to patch}}
\patchcmd{\ALG@doentity}{\item[]\nointerlineskip}{}{}{} 
\renewcommand*\Call[0]{\textproc}

\newcommand{\powerPerChannel}[2]{P^{(#1)}_{#2}}
\newcommand{\psuccPerChannel}[1]{p^{(#1)}_\text{succ}}
\newcommand{\pnodesuccPerChannel}[2]{p^{(#1)}_\text{succ,#2}}

\newcommand{\sigmaAvgPerChannel}[1]{\sigma_{\text{avg}}^{(#1)}}
\newcommand{\RPerChannel}[3]{R^{(#1)}_{#2#3}}
\newcommand{\RoPerChannel}[1]{R_o^{(#1)}}
\newcommand{\accessPerChannel}[2]{\tau^{(#1)}_{#2}}

\newcommand{\throughputCell}[1]{T_{#1}}
\newcommand{\throughputCellCh}[2]{T_{#1}^{(#2)}}

\newcommand{\IPerNode}[2]{I_{#1#2}}
\newcommand{\hPerChannel}[3]{h^{(#1)}_{#2#3}}
\newcommand{\Imaxlimit}[1]{\text{IMAX}_{#1}}

\newcommand{\powerTV}[1]{P^{\text{TV}}_{#1}}
\newcommand{\q}[2]{q_{#1#2}}
\newcommand{\SINR}[2]{\text{SINR}_{#1#2}}

\newcommand{\pidle}{p^{(s)}_\text{idle}}

\newcommand{\Tcol}{T_\text{col}}

\newcommand{\Tnodesucc}[1]{T_\text{succ,#1}}

\newcommand{\g}[2]{g_{#1#2}}

\newcommand{\Itot}[1]{I_{#1}}

\newcommand{\SUB}{\text{channel}}

\newcommand{\TVTXSet}[1]{\text{TV}_{\text{TX}}^{(#1)}}
\newcommand{\TVRXSet}[1]{\text{TV}_{\text{RX}}^{(#1)}}
\newcommand{\AvailC}[1]{S_{#1}}
\newcommand{\AssignC}[1]{\overline{S}_{#1}}
\newcommand{\Channels}{\mathcal{S}}
\newcommand{\ChannelQuality}[2]{\gamma_{#1}^{(#2)}}
\newcommand{\Adj}{[A]}
\newtheorem{lemma}{Lemma}
\usepackage{url}
\usepackage{xcolor}

\newcommand{\SG}[1]{{\color{black} {#1}}}

\usepackage{mathtools}
\begin{document}
%
\title{Optimizing City-Wide \SG{White-Fi} Networks\\ in TV White Spaces}
\author{Sneihil Gopal$^{*}$, Sanjit K. Kaul$^{*}$ and Sumit Roy$^{\dagger}$\\
$^{*}$Wireless Systems Lab, IIIT-Delhi, India,$^{\dagger}$University of Washington, Seattle, WA\\
\{sneihilg, skkaul\}@iiitd.ac.in, sroy@u.washington.edu}

\maketitle
\begin{abstract}
White-Fi refers to WiFi deployed in the TV white spaces. Unlike its ISM band counterparts, White-Fi must obey requirements that protect TV reception. As a result, optimization of citywide White-Fi networks faces the challenges of heterogeneous channel availability and link quality, over location. The former is because, at any location, channels in use by TV networks are not available for use by White-Fi. The latter is because the link quality achievable at a White-Fi receiver is determined by not only its link gain to its transmitter but also by its link gains to TV transmitters and its transmitter's link gains to TV receivers.

In this work, we model the medium access control (MAC) throughput of a White-Fi network. We propose heuristic algorithms to optimize the throughput, given the described heterogeneity. The algorithms assign power, access probability, and channels to nodes in the network, under the constraint that reception at TV receivers is not compromised. We evaluate the efficacy of our approach over example city-wide White-Fi networks deployed over Denver and Columbus (respectively, low and high channel availability) in the USA, and compare with assignments cognizant of heterogeneity to a lesser degree, for example, akin to FCC regulations.
\end{abstract}
\IEEEpeerreviewmaketitle
\section{Introduction}
\label{sec:intro}
TV White Spaces (TVWS) are spectrum licensed for TV broadcast that spectrum regulators, for example, FCC~\cite{fcc} in the US and Ofcom~\cite{ofcom} in \SG{the UK}, have opened for use by unlicensed secondary devices, with approaches prescribed to protect the incumbent TV network from interference by the secondaries. TVWS include bands of $54-698$ MHz~\cite{fcc} in the United States and $470-790$ MHz~\cite{ofcom} in Europe.

\SG{IEEE $802.11$af~\cite{TVstandard} and IEEE $802.22$~\cite{802_22} are the standards developed for wireless networks operating in TV White Spaces. $802.11$af, also called} White-Fi and Super-WiFi, refers to a network of secondary devices that use IEEE $802.11$ (WiFi) like physical layer (PHY) and medium access control (MAC) mechanisms. White-Fi cells may be deployed indoors, with coverage of a few $100$ meters, and outdoors, with coverage as large as $5$ km. Large cells may be used to provide internet access in sparsely populated areas. They may also be desirable when the white space channels available at a location are limited and do not allow for channelization and small cells.
\begin{figure}
\vspace{-5pt}
\begin{center}
\includegraphics[width=0.75\linewidth, bb = 0 0 475 435]{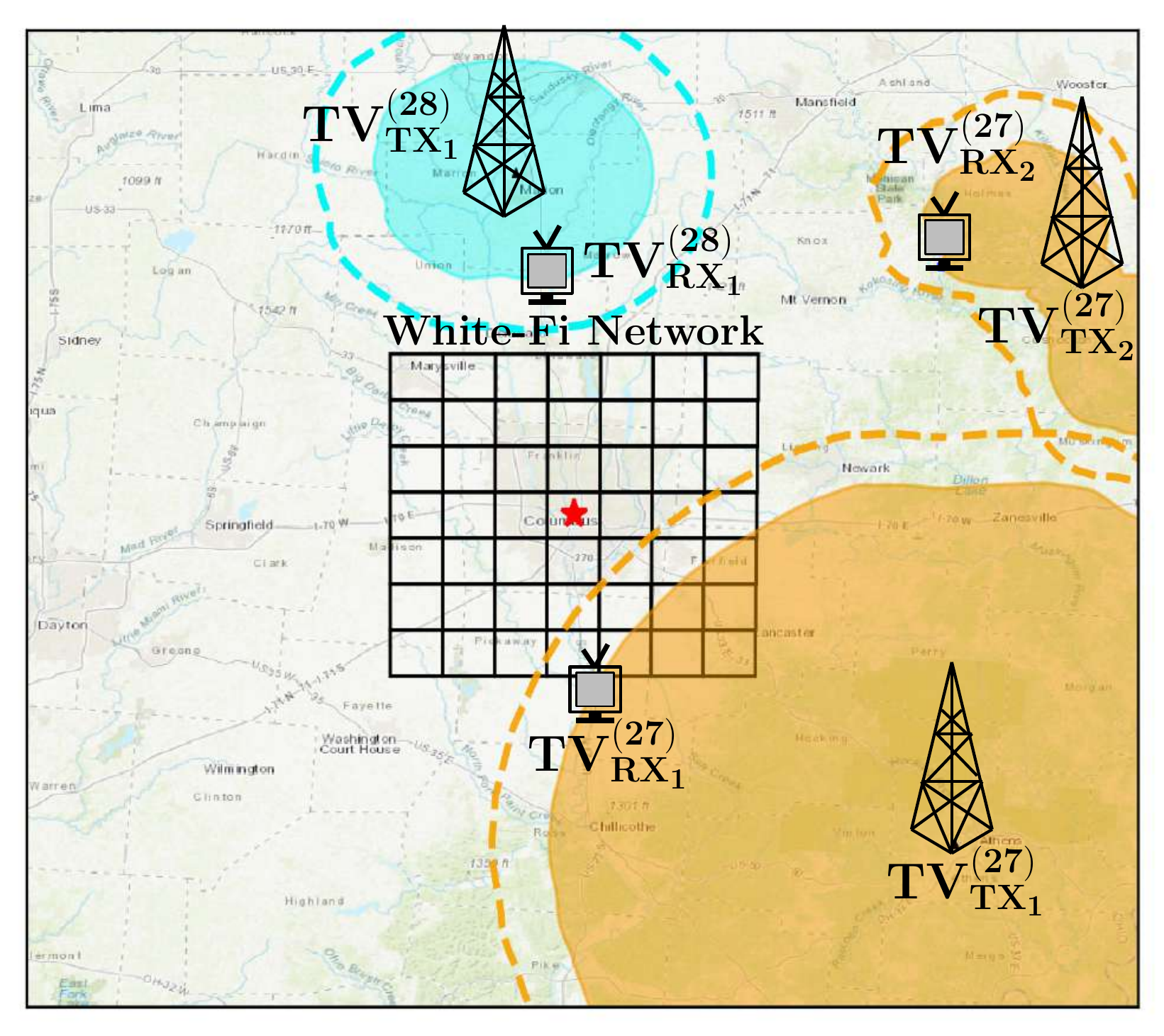}
\end{center}
\vspace{-5pt}
\caption{\small Illustration of a city-wide White-Fi network deployed in Columbus, USA. Three TV networks, operating on channels $27$ and $28$, are shown. For each TV network we show the TV transmitter and a TV receiver. For example, on channel $28$ we have $TV^{(28)}_{TX_1}$ and $TV^{(28)}_{RX_1}$. Around each TV transmitter we show its region of service (solid colored). The boundary of this region is the service contour of the transmitter. We also show the so-called protection contour (dashed line) for each transmitter. It encloses a safety zone in addition to the region of service of the transmitter. The contours were obtained from a database~\cite{sneihil2017} created using information from the FCC.}
\label{fig:network}%
\vspace{-1.5em}
\end{figure}
In this work, we consider a citywide White-Fi network. Figure~\ref{fig:network} provides an illustration. The geographical region covered by the network is tessellated by White-Fi cells. There are also multiple TV transmitters and receivers, with the transmitters servicing locations in and around the White-Fi network. Transmissions due to White-Fi nodes must not impair reception of TV broadcasts at TV receivers. This requirement, as we explain next, leads to fluctuation in \emph{channel availability} and \emph{achievable White-Fi link quality} as a function of location in a White-Fi network, and makes the optimization of such networks distinct from that of traditional WiFi networks operating in the $2.4$ and $5$ GHz unlicensed bands, which see homogeneity in channel availability and achievable link quality. 

\begin{figure*}[t]
\begin{center}
\sbox\twosubbox{%
  \resizebox{\dimexpr.79\linewidth-0.9em}{!}{%
    \includegraphics[height=0.1cm,scale=0.5,bb = 0 0 500 500]{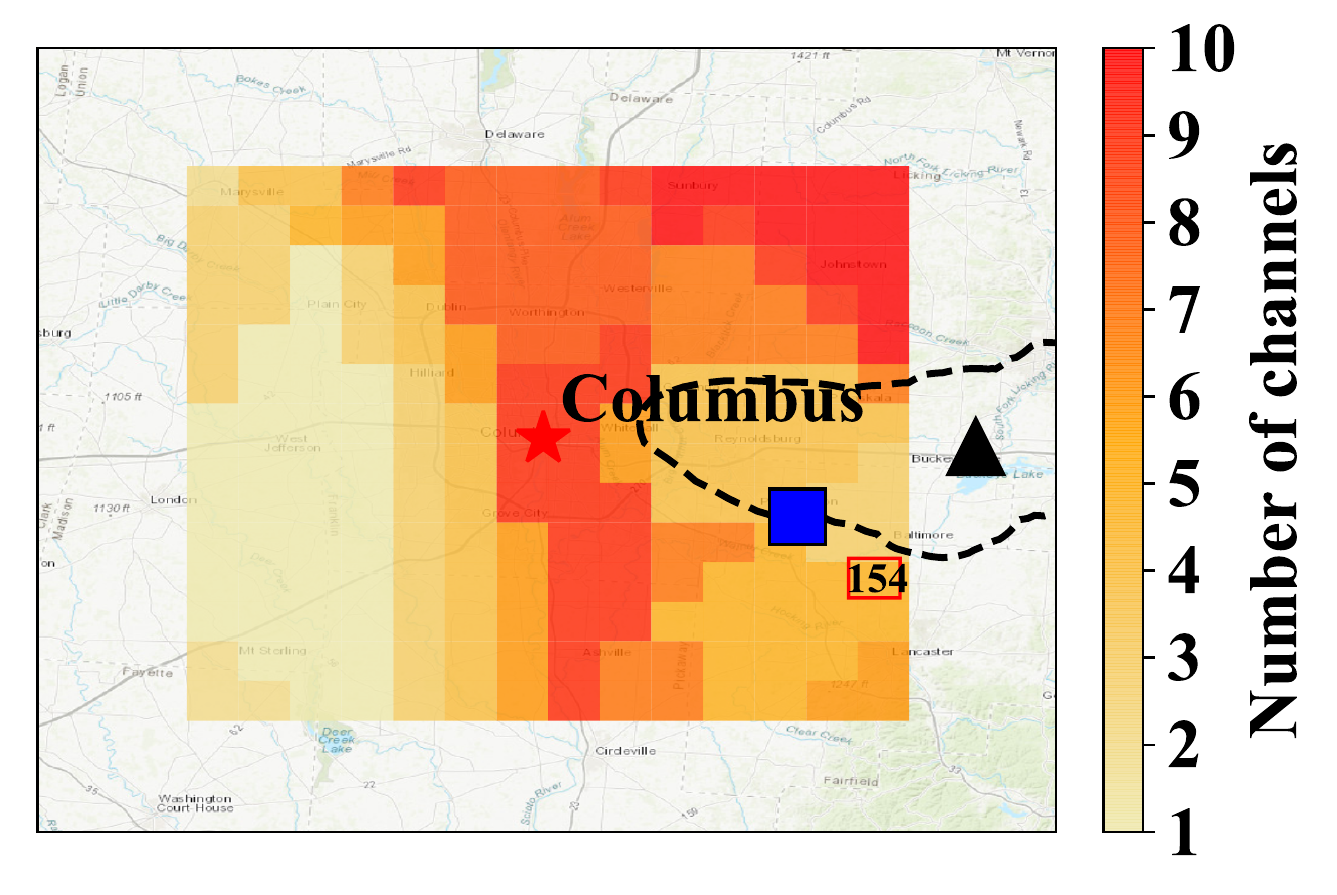}%
    \includegraphics[height=0.1cm,scale=0.5,bb = 0 0 1000 500]{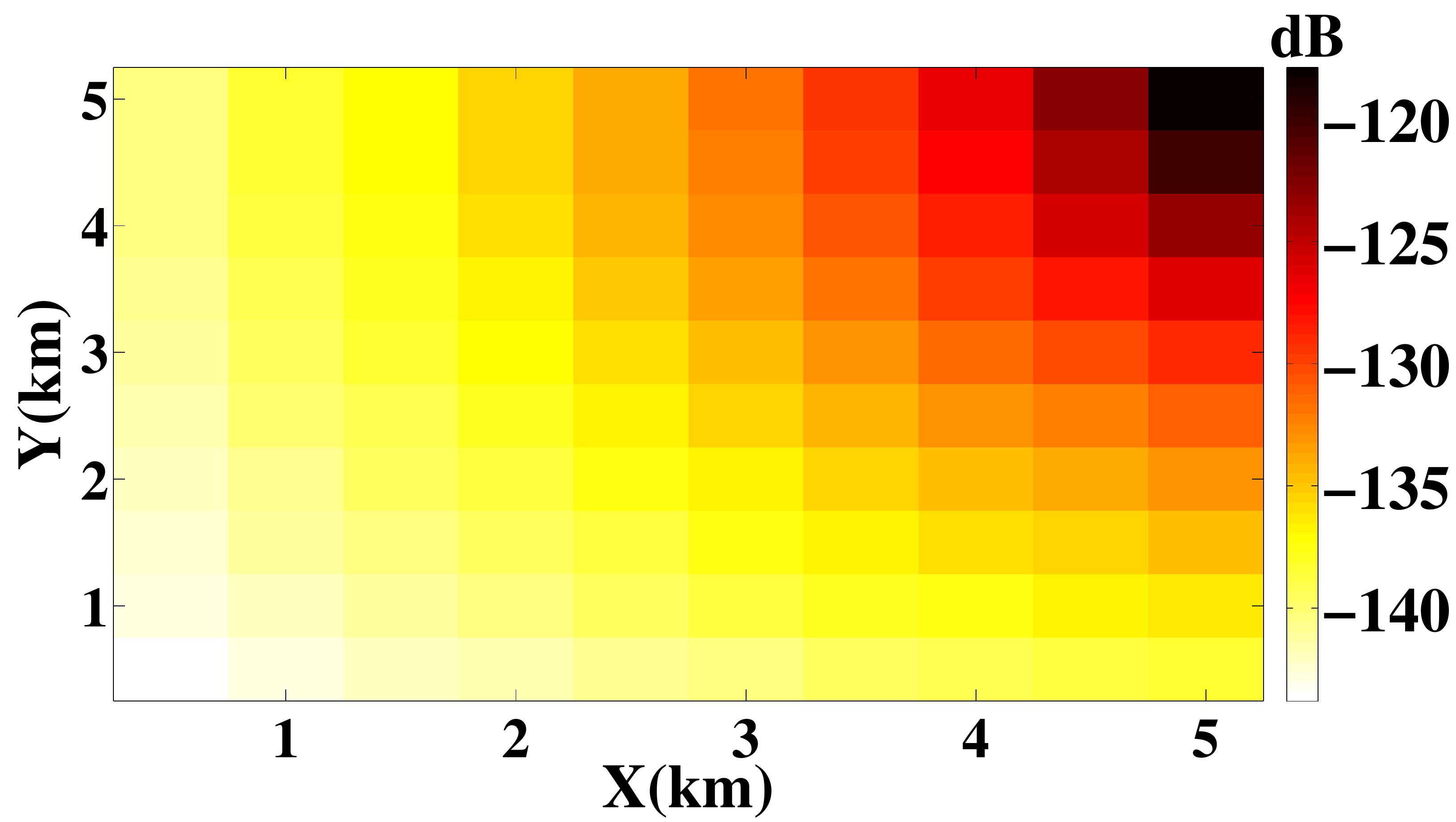}%
    \includegraphics[height=0.1cm,scale=0.5,bb = 0 0 1000 500]{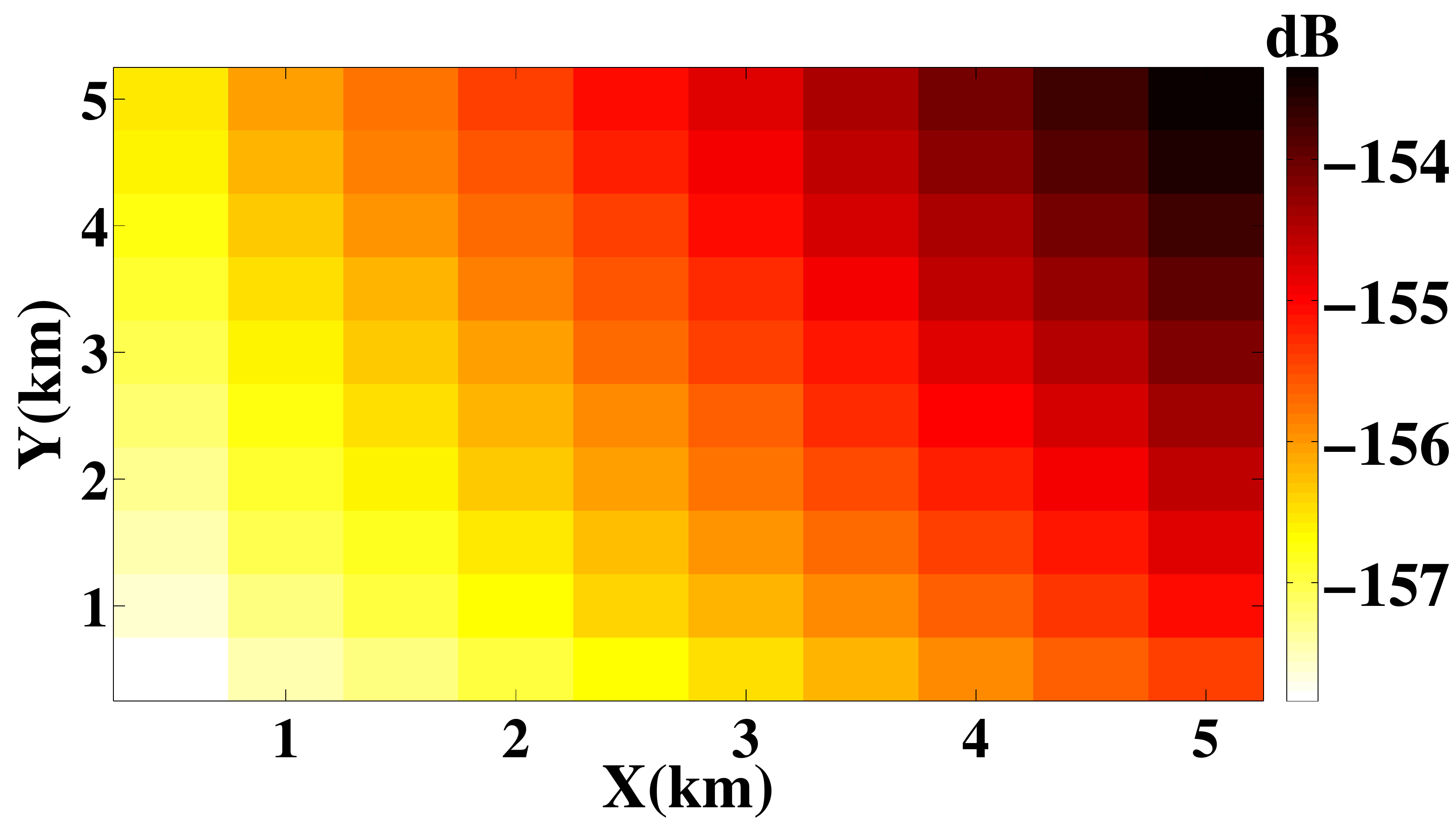}%
  }%
}
\setlength{\twosubht}{\ht\twosubbox}
\centering
\captionsetup[subfloat]{margin=-2.5cm}
\subfloat[]{\label{fig:channelaval}}{%
  \fbox{\includegraphics[height=1.1\twosubht,scale=1]{Figures/Heatmap_channel_availability.pdf}}%
}
\quad
  \captionsetup[subfloat]{margin=-2.65cm}
\subfloat[]{\label{fig:channelgain}}{%
  \fbox{\includegraphics[height=1.1\twosubht,scale=1]{Figures/Channel_gain_variation.pdf}}%
}
\quad
\subfloat[]{\label{fig:channelgainP2S}}{%
  \fbox{\includegraphics[height=1.1\twosubht,scale=1]{Figures/Channel_gain_variation_PrimToSec.pdf}}%
}
\end{center}
\caption{\small{(a) Channel availability in the city of Columbus in the USA over a White-Fi network spread over an area of $4900$ km$^{2}$. Each pixel in the White-Fi network represents a White-Fi cell covering an area of $25$km$^{2}$. Channel availability is computed~\SG{for a scenario where a channel is deemed available for a White-Fi cell if the cell lies outside the service contour of every TV transmitter operating on that channel.} 
(b) Link gains from locations in the White-Fi cell (indexed $154$ in (a)) to the TV receiver (solid blue square in (a)) located about $1$ km away. (c) Link gains to locations in the White-Fi cell $154$ from the TV transmitter (solid black triangle in (a)) located at a distance of $15.7$ km. All link gain calculations assume a path loss model with an exponent of $3$.}}
\label{fig:channel}
\vspace{-1em}
\end{figure*}
\emph{Heterogeneity in channel availability:} The region serviced by a TV transmitter, illustrated in Figure~\ref{fig:network}, is enclosed by its \emph{service contour}. TV transmitters are allocated channels such that they don't create interference within each other's service contours. These channels are known a priori and may be obtained, for example, from~\cite{fccdata}, which has data related to the US. In addition, to protect TV receivers from interference due to transmissions by secondary devices, regulatory bodies disallow a secondary from transmitting over channels being used by TV transmitters if the secondary is within a certain geographic proximity of the TV transmitters. This results in heterogeneous channel availability across locations in a city-wide White-Fi network.

To exemplify, FCC~\cite{fcclicensed} restricts transmissions over the so-called \emph{protected region} around a TV transmitter i.e. a TV channel is available for transmission by secondary nodes at a location only if the location is outside the protected region of each TV transmitter operating on that channel. The protected region, \SG{enclosed by the protection contour (dashed lines in Figure~\ref{fig:network})}, includes the region within the service contour and an additional buffer to protect the TV receivers from secondary interference. Figure~\ref{fig:channelaval} shows the resulting number of channels available, for transmissions by secondary nodes, over an area of $4900$ km$^{2}$ in the city of Columbus, USA. The availability varies over a large range of $1-10$ channels. 

\emph{Heterogeneity in achievable link quality:} The link quality, specifically the signal-to-interference-and-noise-ratio (SINR), of a White-Fi link is impacted by the TV network in a two-fold manner. (a) Link SINR suffers due to interference from TV transmitters operating on the same/interfering channel as that of the link. Specifically, the larger the link gain between a TV transmitter and a White-Fi receiver, the smaller is the SINR of the corresponding White-Fi link. (b) The larger the link gain between a White-Fi transmitter and a TV receiver, the smaller the power the transmitter may use on its link without unduly impacting reception at the TV receiver.

Since White-Fi cells can be large in size, even locations within a cell may see very different link gains to TV transmitters and receivers. Figure~\ref{fig:channelgain} shows the link gains from locations in a White-Fi cell of size $5$ km $\times\ 5$ km to a TV receiver located at a distance of about $1$ km. A spread of about $25$ dB is observed. Figure~\ref{fig:channelgainP2S} shows link gains from a TV transmitter 
located at a distance of about $16$ km from the cell. 
We observe a spread of about $5$ dB in gains. This heterogeneity in gains, within a cell, to and from the TV network does not exist when the nodes are spread over a very small region as is the case in traditional WiFi.

\emph{The optimization problem:} In this work, we investigate optimizing the medium access control (MAC) saturation (every node always has a packet to send) throughput of a citywide White-Fi network. 
Similar to 
traditional WiFi, nodes in a White-Fi cell use the distributed coordination function (DCF)~\cite{bianchi}, which is a carrier sense multiple access and collision avoidance (CSMA/CA) based MAC to gain access to the medium. We want to maximize the MAC saturation throughput of the network under the constraint that reception at TV receivers is not impaired. Specifically, we enforce that the maximum aggregate interference that nodes in the White-Fi network create at any TV receiver is within allowed limits. We optimize over assignments of channel, transmit power, and medium access probability, to nodes in the network. This allows us to adapt to the aforementioned heterogeneities.

\SG{A node may be assigned one or more channels and is capable of operating on multiple channels simultaneously. Specifically, a node attempts access in a channel and maintains DCF state independently of its access attempts and DCF state in its other assigned channels. Also, a node saturates a separate queue for each assigned channel. As detailed later, this allows one to apply the existing DCF model on each channel independently. In practice, such a node may be a software defined radio capable of processing multiple channels at once.

The transmit power allocated to a channel, however, is not independent of the allocation to other channels assigned to it. Specifically, each node in a White-Fi cell has a total power budget which it splits across the assigned channels. This allows for a better allocation of the available power budget, at every node, across assigned channels.}

Unlike our model, FCC regulations propose a simpler model of coexistence with the TV network in which any secondary node outside the protected region can transmit at its full power, as allowed for its device category~\cite{fcc}. 
Despite being conservative, the regulations do not guarantee protection of the TV network from excess \emph{aggregate} interference that results from more than one node transmitting using the allowed power. \SG{Moreover, they result in a wastage of white spaces. These facts were first observed in~\cite{sahai}. The authors showed that limits on aggregate interference at the TV receiver would not only eliminate the possibility of outage at the TV receiver, but also allow secondary nodes to operate anywhere outside the service contour, hence eliminating the need for protected regions and increasing white space availability.}

Unlike the FCC, Ofcom requires secondary transmit power to be a function of location. This requirement is implicit in our constraint on aggregate interference at a TV receiver. The aggregate interference is a function of the transmit powers of the White-Fi nodes and their link gains to the TV receiver, where the gains are a function of the locations of the TV receiver and the White-Fi nodes. In fact, the White-Fi nodes that are assigned the same channel as the TV receiver may be located across different cells. As a consequence, the aggregate interference budget at the TV receiver is shared between nodes in different White-Fi cells. 

There are prior studies such as~\cite{radunovic,hessar1,tan,leResource,shi,leejoint,KTH13} that propose maximization of secondary network throughput under aggregate interference and transmit power budget constraints. However, to the best of our knowledge, our work is the first attempt to model and optimize the DCF throughput of a multi-cell city-wide White-Fi network. 
Our specific contributions are listed next.
\begin{itemize}[leftmargin=*]
\item We formulate the problem of maximizing the DCF saturation throughput of a multi-cell White-Fi network under the constraint that the maximum aggregate interference that White-Fi nodes create at TV receivers is within acceptable limits.
\item We rework the saturation throughput model proposed by Bianchi in~\cite{bianchi} to incorporate per node transmit power and payload rates, per node access probabilities, and an overhead rate used to communicate control packets that is not fixed but results from the maximization.
\item The throughput maximization is a non-linear optimization problem. We propose a two-phase heuristic solution. In the first phase, we assign TV white space channels to cells in the White-Fi network. In the second phase, we assign nodes in the cells their payload transmission rates/transmit powers and access probabilities, over each assigned channel. 
\item We demonstrate the efficacy of the proposed heuristic method over hypothetical deployments of White-Fi networks coexisting with real TV networks in the US cities of Columbus and Denver. Together, these cities provide good examples of heterogeneity in channel availability and link quality in the white spaces. Surprisingly, while Columbus has higher channel availability as compared to Denver, its network throughput is lower.
\item Futher, we compare our approach to a baseline that adheres to restrictions on aggregate interference but allocates the same power and access probabilities to all nodes in a cell, which makes it easier to implement in practice. 
\item Last but not the least, we quantify the reduction in the availability of white spaces that may result from the use of FCC-like regulations when restrictions on aggregate interference from White-Fi nodes must be enforced.
\end{itemize}

Rest of the paper starts with related works in Section~\ref{sec:related}. The network and the saturation throughput model are described in Section~\ref{sec:model}. This is followed by the optimization problem in Section~\ref{sec:opt}. The solution methodology is described in Section~\ref{sec:method}. The hypothetical White-Fi networks and results are respectively in Sections~\ref{sec:simulation} and~\ref{sec:results}. 
We conclude in Section~\ref{sec:conclusions}.

\section{Related Work}
\label{sec:related}
In preliminary work~\cite{sneihilg}, we investigated optimization of DCF throughput of a single White-Fi cell flanked by two TV networks. Authors in~\cite{sahai1,hessar,achtzehn2012,deshmukh} and~\cite{dudda,jantti} have proposed approaches for assessment of TVWS capacity under FCC and ECC regulations, respectively. Contrary to~\cite{sahai1,hessar,achtzehn2012,deshmukh}, the authors in~\cite{sahai} advocate that FCC regulations are stringent and must be replaced by spatially-aware rules for better utilization of TVWS. Motivated by~\cite{sahai}, our work focuses on leveraging heterogeneity in white space availability and link quality to maximize the throughput of a citywide White-Fi network. Also, while authors in~\cite{hessar,achtzehn2012,dudda,deshmukh} provide assessment of TVWS capacity at any location (very short range communication), we are interested in the capacity of an outdoor White-Fi network comprising of long-range links. 

Modelling aggregate interference at a TV receiver from a secondary network has been previously studied in~\cite{alberto1,alberto2,alberto3,amir,tercero,obregon}. In~\cite{koufos1,koufos,KTH12} authors propose an approach for determining permissible transmit powers for secondary networks under aggregate interference constraints. Authors in~\cite{koufos1} quantify the capacity available to a secondary system under constraints on interference at the TV receiver. In~\cite{selen}, authors propose to maximize the sum capacity of a secondary network by setting the power limits for each white space device while limiting the probability of harmful interference created at the primary network. While the aforementioned works,~\cite{alberto1,alberto2,alberto3,amir,tercero,obregon,koufos1,koufos}, study cellular-like secondary systems, in our work, we consider a WiFi-like secondary system.

Similar to our work, authors in~\cite{leejoint,hoang,janku,liang} consider power control and channel allocation in networks operating in TV white spaces. Authors in~\cite{leejoint} consider an overlay secondary system and determine channel allocation and transmit powers according to licensed user activity. Contrary to~\cite{leejoint}, we consider an underlay approach while ensuring the licensed user (TV) is protected. 

Authors in~\cite{hoang} propose an approach that maximizes the throughput of a cellular secondary network while maintaining a required signal-to-interference-and-noise (SINR) for all TV receivers and requires cooperation between secondary devices and TV networks. Authors in~\cite{janku} use the Nash bargaining solution to allocate power and channel to nodes of a secondary network. Their network operates like an infrastructure mode WiFi network (clients communicate via an access point). They don't model WiFi throughput, however. Also, they do not take into consideration the interference to/from the TV networks. Authors in~\cite{liang} propose a channel allocation/power control algorithm to maximize the spectrum utilization of a cellular secondary network while protecting the licensed users and ensuring a minimum SINR for each secondary user. 

Authors in~\cite{KTH13} propose throughput maximization of a WiFi like network in TVWS under aggregate interference. Similar to our work, they allow variable transmit powers. However, they do not model the CSMA/CA based mechanism of the DCF. Instead, they model WiFi link rates to be their Shannon rates. 
Authors in~\cite{wifiXL} propose a white space wide area wireless network that extends WiFi like spectrum sharing to TVWS. Unlike our work, they assume that nodes can transmit at their maximum transmit powers. As a result they do not optimize the interplay between the secondary users and the TV network. 

Authors in~\cite{radunovic} propose a learning algorithm for dynamic rate and channel selection to maximize the throughput of a wireless system in white spaces. Authors in~\cite{hessar1} and~\cite{tan}, propose channel assignment techniques for maximizing throughput of secondary networks in TV white spaces. Similar to our work, authors in~\cite{radunovic,hessar1,tan} propose throughput maximization of secondary networks in TV white spaces, however, they do not consider the impact of aggregate interference from secondary users on TV receivers.~\SG{Other works such as~\cite{802_22_1} and~\cite{802_22_2} study IEEE $802.22$ networks. Authors in~\cite{802_22_1} study channel assignment in $802.22$ networks. Authors in~\cite{802_22_2} study the network coverage, capacity and energy efficiency of LTE networks in TVWS.} 

In summary, while there exist several studies on maximization of secondary network throughput under aggregate interference and power budget constraints, to the best of our knowledge, our work is the first attempt to maximize throughput of a CSMA/CA based citywide White-Fi network in which the secondaries can communicate over long distances.
\section{Network Model}
\label{sec:model}
\begin{figure}
\vspace{-6em}
\begin{center}
\includegraphics[width=0.695\linewidth,bb = 50 0 550 550]{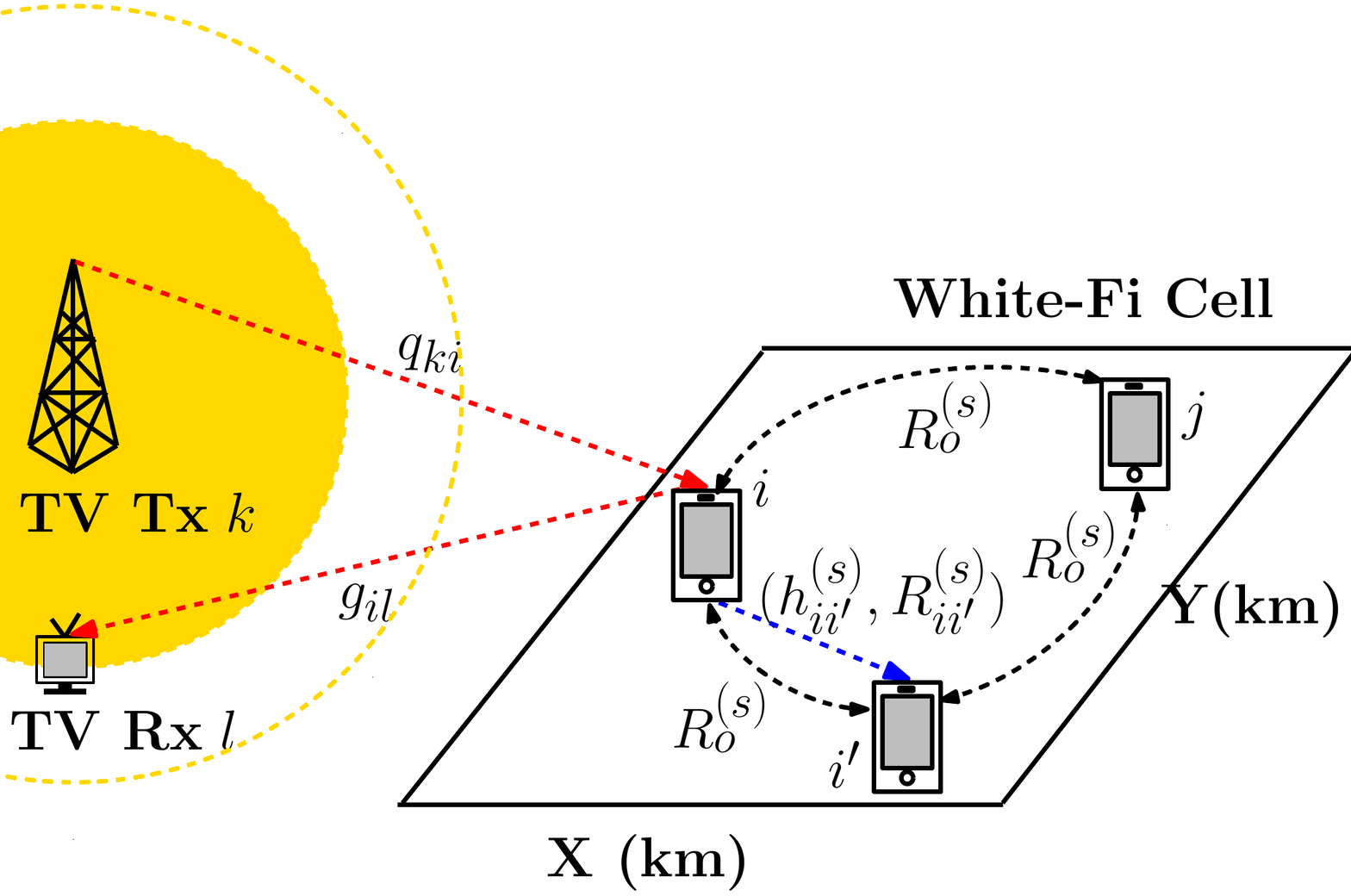}
\end{center}
\vspace{-0.7em}
\caption{\small We summarize the different link gains and rates. A White-Fi cell with nodes $i$, $i'$, and $j$ is shown. The TV network consists of a transmitter (large tower) TV TX $k$ and a receiver TV RX $l$ (assumed to be operating in channel $s$). The gains $\g{i}{l}$ from node $i$ to the TV receiver and $\q{k}{i}$ from the TV transmitter to node $i$, are shown over the corresponding links. Also, we show the link gain  $\hPerChannel{s}{i}{i'}$ between nodes $i$ and $i'$, over the channel $s$. All nodes broadcast control overheads on channel $s$ at rate $\RoPerChannel{s}$ defined in~(\ref{eqn:overheadRate}). Node $i$ sends its data payload to node $i'$ at rate $\RPerChannel{s}{i}{i'}$ defined in~(\ref{eqn:wifiLinkRate}).}
\label{fig:simulationTopology}%
\vspace{-1em}
\end{figure}
Let $\Channels$ be the set of white space channels. Our TV network, illustrated in Figure~\ref{fig:network}, consists of TV transmitters and receivers that operate on channels in $\Channels$. Let $\TVTXSet{s}$ and $\TVRXSet{s}$ respectively be the set of TV transmitters and TV receivers operating on channel $s$. A TV transmitter $k$ has a known transmit power $\powerTV{k}$. Our White-Fi network operates over a geographical region that is tessellated by a set $\mathcal{M}$ of White-Fi cells indexed $1,\ldots, M$. We define $\mathcal{N}$ to be the set of all White-Fi nodes and $\mathcal{N}_m$ to be the set of nodes in cell $m$. It follows that $\mathcal{N} = \cup_{m=1}^M \mathcal{N}_m$. Also, a node must belong to exactly one cell.

White-Fi nodes operating on a channel $s$ will create interference at TV receivers in $\TVRXSet{s}$ and will suffer interference from the TV transmitters in $\TVTXSet{s}$, which are a function of the corresponding link gains. Let $\q{k}{i}$ be the link gain between TV transmitter $k$ and a White-Fi node $i$. Further, let $\g{i}{l}$ be the link gain between node $i$ and TV receiver $l$. These gains are illustrated in Figure~\ref{fig:simulationTopology}. In practice, the knowledge of locations of the TV transmitters, receivers, and the White-Fi nodes, together with a suitable path loss model, can be used to estimate them.

Let $\AvailC{m} \subset \Channels$ be the set of white space channels available in the geographical region covered by cell $m$. Further let $\AssignC{m} \subset \AvailC{m}$ be the channels that are assigned for use in the cell. While $\AvailC{m}$ is known a priori, for example from~\cite{fccdata}, $\AssignC{m}$ is obtained as a result of the proposed throughput optimization. Let $\throughputCellCh{m}{s}$ be the MAC throughput of cell $m$ on a channel $s$ assigned to it. We define the MAC throughput $\throughputCell{m}$ of the cell $m$ as the sum of its throughputs on each assigned channel. The throughput $T$ of the White-Fi network is the sum of the throughputs of its $M$ cells. We have
\begin{subequations}
\begin{align}
\throughputCell{m} &= \sum_{s \in \AssignC{m}} \throughputCellCh{m}{s}\label{eqn:cellThr},\\
T &= \sum_{m=1}^M \throughputCell{m}.\label{eqn:netwkThr}
\end{align}
\end{subequations}
Next we detail the calculation of the throughput $\throughputCellCh{m}{s}$ of cell $m$ on an assigned channel $s$.

\textbf{Throughput of a cell on an assigned channel:} \SG{All White-Fi nodes in a cell (set $\mathcal{N}_m$ for cell $m$) follow the IEEE $802.11$ DCF for medium access in each assigned channel $s$ and independently saturate each assigned channel. The DCF defines a binary-exponential backoff based mechanism that allows nodes to contend for the shared medium. It also involves nodes exchanging control messages including request-to-send (RTS), clear-to-send (CTS), and ACK (acknowledgement), in addition to data payloads. DCF was modeled by Bianchi in~\cite{bianchi} under saturation conditions.}

\SG{Bianchi modeled DCF using a two-dimensional discrete-time Markov Chain with time measured in units of a (variable length) DCF slot. A slot could be an idle slot, or a successful transmission slot, or a collision slot. In an idle slot, no transmission takes place. In a successful transmission slot exactly one node transmits. Lastly, in a collision slot more than one node transmits and this results in all transmitted packets being decoded in error. Saturation throughput is the fraction of an average slot spent in successful payload transmissions.}

\SG{It is worthy of note that Bianchi's model applied to a network of nodes that were sharing one channel. White-Fi nodes, on the other hand, may be assigned multiple channels. However, as detailed in Section~\ref{sec:opt}, channel assignment ensures that adjacent cells are assigned sets of orthogonal channels. As a result, transmissions on a channel in a cell (and the resulting DCF state) are not impacted by transmissions on other cells and channels. Thus, we can model the DCF throughput $\throughputCellCh{m}{s}$ of a cell $m$ on a channel $s$, as if only one channel was being used.}

\SG{To calculate the MAC saturation throughput $\throughputCellCh{m}{s}$, we will use the slot definitions in~\cite{bianchi}. However, we will rewrite the slot lengths to incorporate various link gains and rates (summarized in Figure~\ref{fig:simulationTopology}) that are intrinsic to our problem. We will also rewrite the probabilities of occurrence of the different slots to incorporate the fact that each White-Fi node in cell $m$ could use a different access probability in every assigned channel $s$.} 

\emph{On slot lengths and access probabilities:} A node in a cell will transmit data payloads over all channels assigned to the cell. Let $\powerPerChannel{s}{i}$ be the power with which such a node $i$ transmits its unicast data payload over channel $s$. Without loss of generality, assume that a certain other node $i'$ in the cell is the destination for $i$'s payload. Let the gain of the link between $i$ and $i'$ over channel $s$ be $\hPerChannel{s}{i}{i'}$ (see Figure~\ref{fig:simulationTopology}). Note that $i$ and $i'$ can communicate using all channels in the set $\AssignC{m}$ and the link gain $\hPerChannel{s}{i}{i'}$ is a function of the channel $s\in \AssignC{m}$ under consideration. The payload rate $\RPerChannel{s}{i}{i'}$ that may be achieved between $i$ and $i'$, for a channel bandwidth of $B$ Hz and thermal noise of spectral intensity $N_0$ Watts/Hz, is the Shannon rate of the link and is given by 
\begin{align}
&\RPerChannel{s}{i}{i'} = B\log_2(1 + \SINR{i}{i'})\ \text{bits/sec},\label{eqn:wifiLinkRate}\\
&\text{where}\ \SINR{i}{i'} = \frac{\hPerChannel{s}{i}{i'} \powerPerChannel{s}{i}}{BN_0 + \sum_{k\in \TVTXSet{s}} \q{k}{i'} \powerTV{k}}.\label{eqn:wifiLinkSINR}
\end{align}
The numerator of $\SINR{i}{i'}$ is the power received at node $i'$ from $i$. The denominator is the sum of receiver noise power at $i'$ and the sum of interference powers received from TV transmitters on the channel $s$. For any TV transmitter $k$ on the channel, the interference power is the product of its transmit power $\powerTV{k}$ and its link gain $\q{k}{i'}$ to the White-Fi receiver $i'$. Node $i$ achieves the payload rate $\RPerChannel{s}{i}{i'}$ only for the fraction of time it gets successful access to the medium.

We will enforce that all nodes in a cell can decode messages that help regulate access to the medium, for example, RTS, CTS, and ACK, sent from any other node in the cell. We will refer to such messages as overheads. So while a data payload sent by $i$ to $i'$ at rate $\RPerChannel{s}{i}{i'}$ may not be correctly decoded by a node other than $i'$, all overhead messages sent by $i$ must be correctly decoded by all nodes.

These overheads transmitted by node $i$ can be correctly decoded by any other node in the cell if they are sent at a rate not greater than $B \log_2(1 + \min_{j\in \mathcal{N}_m, j\ne i} \SINR{i}{j})$. In this work, for simplicity of exposition, we will assume that all nodes within a cell $m$, use the same rate $\RoPerChannel{s}$ to send overheads. Note that this underestimates the achievable throughput of the cell. The overhead rate is given by
\begin{align}
\RoPerChannel{s} = \min_{i\in\mathcal{N}_m} B \log_2(1 + \min_{\substack{j\ne i\\j\in\mathcal{N}_m}} \SINR{i}{j}).\label{eqn:overheadRate}
\end{align}

Let $\accessPerChannel{s}{i}$ be the steady state probability with which White-Fi node $i$ accesses the wireless medium during a DCF slot for transmitting its payload, over channel $s$. The probability $\pnodesuccPerChannel{s}{i}$ that a transmission by $i$ is successful and the probability $\pidle$ that a slot is idle are, respectively,
\begin{align}
\hspace{-.5em}\pnodesuccPerChannel{s}{i} = \accessPerChannel{s}{i}\prod\limits_{\substack{j\in \mathcal{N}_m \\ j\neq i}}(1-\accessPerChannel{s}{j})\ \text{and}\ \pidle = \prod_{j\in \mathcal{N}_m}(1-\accessPerChannel{s}{j}).
\label{eqn:psuccAndIdleForANode}
\end{align}
Let idle slots be of duration $\sigma$. In practice, this is specified by the $802.11$ standard. Let $L$ bits be the size of payload in a packet transmitted by any White-Fi node $i$. A slot that sees a successful transmission consists of the payload, overhead bits including packet headers, RTS/CTS, ACK, and overheads due to inter frame spacings like DIFS. Let $O_{\text{bits}}$ be the number of overhead bits. They are transmitted at the overhead rate $\RoPerChannel{s}$ defined in~(\ref{eqn:overheadRate}). The payload is transmitted by node $i$ to its destination node $i'$ at rate $\RPerChannel{s}{i}{i'}$ defined in~(\ref{eqn:wifiLinkRate}). Let $O_{\text{sec}}$ denote the frame spacing related overheads. Therefore, the total duration of a successful transmission slot of node $i$ is given by $\Tnodesucc{i} = O_{\text{sec}} + O_{\text{bits}}/\RoPerChannel{s} + L/\RPerChannel{s}{i}{i'}$.

Finally, a slot that sees a collision has $L_{\text{col}}$ bits and $L_{\text{colsec}}$ time overheads. The duration of a collision is given by $\Tcol = L_{\text{col}}/\RoPerChannel{s} + L_{\text{colsec}}$. On use of RTS/CTS, which we assume in this work\footnote{The case when the network does not use RTS/CTS introduces variable length collisions, where the length is a function of the payload rates of the colliding transmissions. While, this can be incorporated in the model, the resulting expression for the length of the collision slot becomes unwieldy. Also, as is shown in~\cite{bianchi}, under saturation conditions the maximum throughput with or without RTS/CTS is the same. In fact, using RTS/CTS is more desirable as it makes the throughput less sensitive to small changes in access probability.}, 
only RTS packets may collide. As a result, $\Tcol$ is the same irrespective of which nodes' transmissions were involved in a collision. The length $\sigmaAvgPerChannel{s}$ of an average DCF slot in our cell, over channel $s$, is
\begin{align}
&\sigmaAvgPerChannel{s} = \pidle \sigma +  \sum_{i\in \mathcal{N}_m}\pnodesuccPerChannel{s}{i} \Tnodesucc{i} + (1 - \pidle - \psuccPerChannel{s}) \Tcol,
\label{eqn:avgSlot}
\end{align}
where $\psuccPerChannel{s} = \sum_{i\in\mathcal{N}_m} \pnodesuccPerChannel{s}{i}$. Note that $\psuccPerChannel{s}$ is simply the probability that a DCF slot sees a successful transmission. We have an average of $\psuccPerChannel{s} L$ payload bits transmitted successfully in the network over an average slot of length $\sigmaAvgPerChannel{s}$. The throughput $\throughputCellCh{m}{s}$ bits/sec can thus be obtained as
\begin{align}
\throughputCellCh{m}{s} =  {\psuccPerChannel{s} L}/{\sigmaAvgPerChannel{s}}.
\label{eqn:perCellPerChannelTh}
\end{align}

\emph{Comments on the DCF model:} \SG{Given traffic sources in practice, networks usually do not operate in saturation conditions. However, saturation throughput is often studied for the following reasons: (a) It indicates the steady state throughput that is obtainable when the network is heavily loaded. Also, while the non-saturated maximum throughput may be greater and may seem like a more useful characterization, unlike saturation throughput, it is not feasible to have a network operating stably at the maximum~\cite{bianchi}~\cite{bertsekas}. (b) Saturation simplifies analysis by allowing one to avoid modeling traffic and concentrate entirely on the DCF mechanism. That said, earlier work~\cite{duffy} has attempted extending Bianchi's model to the non-saturated case by approximating non-saturated traffic flows by defining a probability that a packet is present at a node. The throughput of a White-Fi network may be derived, in a manner similar to the saturated case (probability is $1$), to incorporate such an approximation of traffic.}

Lastly, the model proposed by Bianchi~\cite{bianchi} assumes that all nodes can decode control messages such as request-to-send (RTS), clear-to-send (CTS), and ACK, sent from any other node in the cell. 
This precludes the possibility of hidden nodes in the network. There are many works that extend~\cite{bianchi} to model the throughput in the presence of hidden nodes. It is noteworthy that the extension in~\cite{ekici_ieee_2008} uses the same basic form of throughput as in~\cite{bianchi} and is amenable to per node payload rates and access probabilities, and optimizable overhead rates. That said, the extension to hidden nodes is non-trivial. Among other things, not all nodes in the cell see the same set of hidden nodes. Importantly, the basic interplay between the White-Fi network and the primary, which involves interference created by TV transmitters at White-Fi nodes and the interference created by the nodes at TV receivers, and the impact of interference on payload and overhead rates in the network, is captured well by the White-Fi network model we consider in this work.
\section{Optimization Problem}
\label{sec:opt}
We want to optimize the network throughput $T$, defined in~(\ref{eqn:netwkThr}). However, transmissions by nodes in the White-Fi network must not impair TV reception. Specifically, we will impose limits on the aggregate interference that White-Fi nodes may create at TV receivers. In addition, we also impose (a) adjacent White-Fi cells must be assigned orthogonal channels, (b) a finite power budget per node in the White-Fi network, and (c) time fairness amongst White-Fi links in a cell.

\emph{Limits on aggregate interference:} Let $l$ be a receiver in the set $\TVRXSet{s}$. A White-Fi node $i$ transmits in a DCF slot, over channel $s$, with probability $\accessPerChannel{s}{i}$. As a result, the interference $\IPerNode{i}{l}$ that the node $i$ creates at $l$ in any DCF slot is a Bernoulli random variable
\begin{align}
\IPerNode{i}{l} = 
\begin{cases}
\g{i}{l} \powerPerChannel{s}{i} & \text{w.p. } \accessPerChannel{s}{i},\\
0 & \text{otherwise}.
\end{cases}
\label{eqn:int}
\end{align}
Recall that $\g{i}{l}$ is the link gain between node $i$ and TV receiver $l$ and $\powerPerChannel{s}{i}$ is the power used by $i$ on channel $s$. The receiver $l$ will suffer interference from all nodes that are transmitting over channel $s$. These are the nodes that belong to cells that have been assigned the channel $s$. The aggregate interference $\Itot{l}$ at $l$ is
\begin{align}
\Itot{l} = \sum_{m: s\in \AssignC{m}} \sum_{i\in \mathcal{N}_m} \IPerNode{i}{l}.
\end{align}

In our earlier work on a single White-Fi cell~\cite{sneihilg}, we had considered, separately, limits on the maximum of the random variable $\Itot{l}$ and its expectation $E[\Itot{l}]$. Since the resulting qualitative insights were similar, in this work, we restrict our investigation to limits on the maximum of $\Itot{l}$. Let the desired limit on maximum aggregate interference at TV receiver $l$ be $\Imaxlimit{l}$. We require that the maximum aggregate interference created by White-Fi nodes at any TV receiver $l$, operating on a channel $s$ assigned to any cell in the White-Fi network, not exceed $\Imaxlimit{l}$. Observe from~(\ref{eqn:int}) that $\max(\IPerNode{i}{l}) = \g{i}{l}\powerPerChannel{s}{i}$. Our desired constraint is given by the following system of inequalities, one for each TV receiver.
\begin{align}
\max(\Itot{l}) = \sum_{m: s\in \AssignC{m}} \sum_{i\in \mathcal{N}_m} \g{i}{l} \powerPerChannel{s}{i} \le \Imaxlimit{l},\nonumber\\ \forall l \in \TVRXSet{s}, \forall s \in \cup_{m=1}^M \AssignC{m}.
\label{eqn:maxConst}
\end{align}

Observe that the aggregate interference at a TV receiver results from nodes in one or more White-Fi cells that are assigned the white space channel on which the TV receiver operates. As a result, nodes in a cell cannot be assigned transmit powers on a given channel independently of nodes in other cells that have been assigned the same channel.

\emph{Channel assignment constraint:} Let the $M\times M$ matrix $\Adj = \{a_{ij}\}$ be the adjacency matrix of the cells in the White-Fi network. For any two cells $m_1,m_2 \in \mathcal{M}$, we have $a_{m_1m_2} = 1$ if $m_1$ and $m_2$ are adjacent\footnote{Any two cells $m_1,m_2 \in \mathcal{M}$ are adjacent if the distance between the cells is $\leq d$, where, $d$ is the distance at which the received power is $3$ dB less than the noise floor. For a path loss exponent of $3$, $d = 2.59$ km.}, otherwise, $a_{m_1m_2} = 0$. Nodes in cells that are adjacent can interfere with each other's transmissions. 
We require adjacent cells to be assigned orthogonal channels.
\begin{align}
\AssignC{m_1} \cap \AssignC{m_2} = \phi  \qquad \forall m_1, m_2 \text{ s.t. }a_{m_1m_2} = 1.
\label{eqn:adjacency}
\end{align}

\emph{Power budget constraint:} We assume that every White-Fi node has a total power budget of $P_T$. The node may split this power over one or more assigned channels. We require 
\begin{align}
\sum_{s\in \AssignC{m}} \powerPerChannel{s}{i} \le P_T\qquad \forall i\in\mathcal{N}_m, \forall m \in \mathcal{M}.
\label{eqn:sumPower}
\end{align}

\emph{Time fairness constraint:} Finally, we enforce time fairness across links on each $\SUB$ within a cell. That is, on an average, every link $i$ in a cell spends the same fraction of time transmitting a payload successfully to its destination on an assigned $\SUB$ $s$. This constraint is essential to ensure that links with high link quality don't dominate access to the medium. Recall that the payload is $L$ bits. The average fraction of time spent by a node $i$ transmitting its payload successfully to a node $i'$ over an average slot $\sigmaAvgPerChannel{s}$ is given by $(\pnodesuccPerChannel{s}{i} L/\RPerChannel{s}{i}{i'})/\sigmaAvgPerChannel{s}$.
Given another node $j$ transmitting to $j'$, to satisfy time fairness, we must satisfy the conditions
\begin{align}
\frac{(1 - \accessPerChannel{s}{i})}{\accessPerChannel{s}{i}} \RPerChannel{s}{i}{i'} &= \frac{(1 - \accessPerChannel{s}{j})}{\accessPerChannel{s}{j}} \RPerChannel{s}{j}{j'},\nonumber \\ &\forall i,i',j,j'\in \mathcal{N}_m, \forall s\in \AssignC{m}, \forall m\in \mathcal{M}.
\label{eqn:fairness}
\end{align}

We want to maximize the throughput $T$ of the White-Fi network, which is given by equation~(\ref{eqn:netwkThr}), under the above defined constraints. Our optimization problem is
\begin{align}
\text{Maximize:}&\ T,\quad {\text{subject to:}}\  (\ref{eqn:maxConst}),(\ref{eqn:adjacency}),(\ref{eqn:sumPower}),(\ref{eqn:fairness}).\label{eqn:opt-limit-max}
\end{align}
Our variables of optimization are the sets $\AssignC{m}$ of channels assigned to cells $m\in \mathcal{M}$, the powers $\powerPerChannel{s}{i}$ and medium access probabilities $\accessPerChannel{s}{i}$ assigned to any node $i$ on any channel $s$ that is assigned to the node's cell. This throughput maximization is a non-linear optimization problem that is non-convex in the variables.
\section{Solution Methodology}
\label{sec:method}
We propose a heuristic method that carries out \emph{Channel Assignment} followed by \emph{Power and Access Probability Assignment}.
\begin{itemize}
\item \textit{Channel Assignment:} For each White-Fi cell, assign a set of channels from those available to the cell such that adjacent cells are assigned orthogonal channels (constraint Equation~(\ref{eqn:adjacency})).
\item \textit{Power and Access Probability Assignment:} Assign transmit power and access probability to every node in the White-Fi network, for each channel assigned to it, such that constraints~(\ref{eqn:maxConst}),~(\ref{eqn:sumPower}), and~(\ref{eqn:fairness}) are satisfied.
\end{itemize}
\subsection{Channel Assignment}
\begin{algorithm}[t]
\small
  \caption{Channel Assignment}
  \label{alg:channelassign}
  \begin{algorithmic}[1]
  \Require $\mathcal{M},\Adj,\AvailC{m},\forall m \in \mathcal{M}$; 
  \Ensure $\AssignC{m}, \forall m \in \mathcal{M}$;\\
	\textbf{Set} $m_1,m_2,\ldots,m_{M}$ such that $d(m_1) \le d(m_2) \ldots \le d(m_{M})$, where $m_1,\ldots,m_{M} \in \mathcal{M}$;
  	\While {$\bigcup_{m \in \mathcal{M}} \AvailC{m} \neq \phi$}\label{alg:channelassign0}
		\For{$m = m_1,\ldots,m_{M}$}\label{alg:channelassign1}
			\If {$\AvailC{m} \neq \phi$} 
			   \State $s^* \gets$ $\arg\max_{s \in \AvailC{m}} \gamma_{s}$;
			   \State $\AssignC{m} \gets \AssignC{m} \cup s^*$;\enspace\Comment{assign selected channel to cell}
			   \State $\AvailC{m} \gets \AvailC{m}\setminus\{s^*\}$;\enspace\Comment{remove assigned channel from list of available channels}
			   \State \Call{UpdateChannelList}($s^*,m$);
 		  \EndIf
		\EndFor \label{alg:channelassign2}   
	\EndWhile
	
  \Function{UpdateChannelList}{$s,m$}
   \State $A_m \gets$ $\{m'\in\mathcal{M}:a_{mm'} = 1\}$; \Comment{get neighboring cells of cell $m$}
   \State $A'_m \gets \{m' \in A_{m}:s\in \AvailC{m'}\}$; \Comment{get those neighboring cells with channel $s$ in available channel list}
   \State $\AvailC{m'} \gets \AvailC{m'}\setminus \{s\}$, $\forall m'\in A'_m$;
  \EndFunction
  \end{algorithmic}
\end{algorithm}
The channel assignment problem can be modeled as a \emph{graph coloring problem}~\cite{wanglist}. We abstract the White-Fi network as an undirected graph $G=(V,E)$ with set $V$ of vertices and $E$ of edges. The White-Fi cells in the network are the vertices of the graph and an edge exists between any two adjacent cells. We have, the set of vertices $V=\mathcal{M}$. Also, if $a_{m_1m_2} = 1$ for cells $m_1, m_2 \in \mathcal{M}$, then an edge between them is in set $E$. White-Fi channels (colors) must be assigned to the cells such that no two adjacent cells are assigned the same channel (color). 

We would like to exploit the fact that the presence of the TV network causes the link quality that may be achieved by nodes in cell $m$ to differ over the set $\AvailC{m}$ of available channels. To this end, we quantify the achievable link quality when using channel $s$ available in cell $m$ as 
\begin{align}
\ChannelQuality{m}{s} = \min_{i\in \mathcal{N}_m,l\in \TVRXSet{s}} \frac{\Imaxlimit{l}/\g{i}{l}}{(BN_0 + \sum_{k\in \TVTXSet{s}} \q{k}{i} \powerTV{k})}.
\label{eqn:qualitymetric}
\end{align}
Note that the numerator $\Imaxlimit{l}/\g{i}{l}$ is the maximum transmit power that node $i$ in cell $m$ can use without exceeding the limit $\Imaxlimit{l}$ on interference at $l$. Typically, since other nodes may transmit over the channel, the transmit power that $i$ will be able to use will be smaller. The denominator consists of the sum of interference powers received from all TV transmitters at node $i$ together with the receiver noise at $i$. Thus the ratio is the maximum SINR that node $i$ can achieve in channel $s$, given TV receiver $l$. The channel quality $\ChannelQuality{m}{s}$ is therefore the smallest SINR achieved by any node on channel $s$ in cell $m$, over all $l\in \TVRXSet{s}$.

Solving the graph coloring problem optimally is known to be NP-complete~\cite{graphNP}. \SG{We propose a heuristic approach that is summarized in Algorithm~\ref{alg:channelassign}. The algorithm takes as input the set of cells $\mathcal{M}$, the adjacency matrix $\Adj$, and the set of available channels $\AvailC{m}$ for each cell $m$. It returns the set of assigned channels for each cell. Channel assignment is performed in multiple rounds. In every round (lines~\ref{alg:channelassign1}-\ref{alg:channelassign2}), we assign channels to cells in ascending order of their degree\footnote{This method is similar to that in~\cite{pengutilization}, in which the nodes obtain different \emph{rewards} on different channels, and would like to choose channels to optimize their rewards.}. The degree $d(m)$ of cell $m$ is the number of cells adjacent to it. That is $d(m) = \sum_{j\in \mathcal{M}, j\ne m} a_{mj}$. In a round, cell $m$ is assigned a channel that has the largest $\ChannelQuality{m}{s}$ in the set $\AvailC{m}$ of channels available in $m$. The chosen channel $s^*$ is added to the set of assigned channels $\AssignC{m}$ and removed from $\AvailC{m}$. It is also removed from the sets of available channels of all adjacent cells of $m$ (function \textsc{UpdateChannelList} in Algorithm~\ref{alg:channelassign}). The algorithm repeats the above in the next round in case there is at least one available channel in any cell in the network (see condition in line~\ref{alg:channelassign0}).}
%
\subsection{Power and Access Probability Assignment}
\begin{algorithm}[t]
\small
 \caption{\small Power and Access Probability Assignment}
 \label{alg:solve}
  \begin{algorithmic}[1]
  \Ensure $\mathbf{\Psi}^*, \mathbf{P}^*$;
  \State $iter\gets 0$; \State $\text{Throughput}(iter) \gets 0$;
  \State $\mathbf{P}_0 \gets\ $\emph{Power Initialization}; \Comment{Problem~(\ref{eqn:init1})-(\ref{eqn:init2})}
  \State $\mathbf{P}^* \gets \mathbf{P}_0$;
  \While {true}
  \State $iter \gets iter + 1$;
  \State $\mathbf{\Psi}^* \gets\ $\emph{Solve for Access} $(\mathbf{P}^*)$;\Comment{Problem~(\ref{eqn:solveForAccess})}
  \State $\mathbf{P}^* \gets\ $\emph{Solve for Power} $(\mathbf{\Psi}^*)$;\Comment{Problem~(\ref{eqn:solveForPower})}
  \State $\text{Throughput}(iter) \gets T(\mathbf{\Psi}^*,\mathbf{P}^*)$;
  \If {$|\text{Throughput}(iter) - \text{Throughput}(iter-1)| < \epsilon$}
  \State break;
  \EndIf  
  \EndWhile
  \end{algorithmic}
  \vspace{-0.05in}
\end{algorithm}
For every assigned channel $s$, we now assign transmit powers $\powerPerChannel{s}{i}$ and access probabilities $\accessPerChannel{s}{i}$ to all nodes $i$ in cells that were assigned $s$. Let the vector of transmit powers and access probabilities be $\mathbf{P}$ and $\mathbf{\Psi}$, respectively. The elements of $\mathbf{P}$ are $\powerPerChannel{s}{i}$, for any node $i$ in $\mathcal{N}$ and channel $s$ in $\Channels$. Likewise, the elements of $\mathbf{\Psi}$ are the $\accessPerChannel{s}{i}$. If channel $s$ is not assigned to a cell $m$, then for all nodes $i$ in the cell $\powerPerChannel{s}{i} = 0$ and $\accessPerChannel{s}{i} = 0$. Further, $T(\mathbf{\Psi},\mathbf{P})$ is the throughput~(\ref{eqn:netwkThr}) of the network when the access probability and transmit power vectors are $\mathbf{\Psi}$ and $\mathbf{P}$, respectively. 

We split this assignment problem into the sub-problems of \emph{Power Initialization}, \emph{Solve for Access}, and \emph{Solve for Power}. \emph{Power Initialization} gives us an initial power allocation $\mathbf{P}_0$ that satisfies~(\ref{eqn:maxConst}) and~(\ref{eqn:sumPower}). \emph{Solve for Access} finds the $\mathbf{\Psi}$ that solves the throughput optimization problem~(\ref{eqn:opt-limit-max}), for a power allocation $\mathbf{P}^*$ that satisfies~(\ref{eqn:maxConst}) and~(\ref{eqn:sumPower}). \emph{Solve for Power} finds a power allocation $\mathbf{P}$ that solves~(\ref{eqn:opt-limit-max}), for an access probability assignment $\mathbf{\Psi}^*$ obtained from \emph{Solve for Access}. Having found an initial power assignment, we iterate over \emph{Solve for Access}, and \emph{Solve for Power} till the obtained throughput~(\ref{eqn:netwkThr}) is judged (empirically) to have converged. Algorithm~\ref{alg:solve} summarizes the approach. We next describe the sub-problems.

\subsubsection{Power Initialization}
We want to initialize the vector of transmit powers such that~(\ref{eqn:maxConst}) and~(\ref{eqn:sumPower}) are satisfied. We formulate a simplified problem to do the same. We proceed by assuming that nodes in a cell take turns to transmit their payloads. Further, during its turn a node $i$ in cell $m$ transmits the data payload of $L$ bits \emph{simultaneously using all} assigned channels. The resulting payload rate is $\sum_{s \in \AssignC{m}}\RPerChannel{s}{i}{i'}$. It also transmits header and other overhead information at a rate $\overline{R}$ that is the smallest rate between any two nodes in the cell. For cell $m$, we have $\overline{R} = \min_{s\in \AssignC{m},i,i'\in\mathcal{N}_m} \RPerChannel{s}{i}{i'}$, where $\RPerChannel{s}{i}{i'}$ was defined in~(\ref{eqn:wifiLinkRate}). We will include all the time ($O_\text{sec}$) and bit ($O_{\text{bit}}$) overheads that were included for a node when calculating the White-Fi cell throughput $\throughputCell{m}$ defined in~(\ref{eqn:cellThr}). The time $t_i$ taken by node $i$'s unicast transmission (payload and overheads) to $i'$ is $t_i = {L}{\left(\sum\limits_{s \in \AssignC{m}}\RPerChannel{s}{i}{i'}\right)}^{-1} + \frac{O_{bits}}{\overline{R}} + O_{sec}.$ 
Node $i$ transmits a payload of $L$ over a time of $\sum_{i\in \mathcal{N}_m} t_i$, which is the time that is required for all nodes in $m$ to take their turn. Thus, the throughput of node $i$ is $L/\sum_{i\in \mathcal{N}_m} t_i$ and that of cell $m$ is $|\mathcal{N}_m| L/\sum_{i\in \mathcal{N}_m} t_i$. 

The network throughput, obtained by summing over all cells, is $\sum_{m \in \mathcal{M}}|\mathcal{N}_m|L/\sum_{i\in \mathcal{N}_m} t_i$. We want to solve for the power vector that maximizes the network throughput under the sum power constraint given by~(\ref{eqn:sumPower}), and the constraint~(\ref{eqn:maxConst}) that limits the maximum aggregate interference. The resulting convex optimization problem is given by
\begin{align}
&\text{Maximize:}\enspace \sum_{m \in \mathcal{M}}|\mathcal{N}_m| L \left(\sum_{i\in\mathcal{N}_m} t_i\right)^{-1},\label{eqn:init1}\\
&{\text{subject to:}}\enspace(\ref{eqn:maxConst}),~(\ref{eqn:sumPower}).\label{eqn:init2}
\end{align}
The optimizer is the initial estimate $\mathbf{P}_0$.

\subsubsection{Solve for Access}
We solve for a vector of access probabilities $\mathbf{\Psi}$ that maximizes the White-Fi network throughput $T$ defined in~(\ref{eqn:netwkThr}), for a given transmit power vector $\mathbf{P}^*$ that is obtained from either \emph{Power Initialization} or \emph{Solve for Power}. Such power vectors satisfy constraints~(\ref{eqn:maxConst}) and~(\ref{eqn:sumPower}) for any $\mathbf{\Psi}$. However, the time fairness constraint~(\ref{eqn:fairness}) must be enforced. The optimization problem is
\begin{align}
\text{Maximize:}&\ T(\mathbf{\Psi},\mathbf{P}^*),\quad {\text{subject to:}}\ (\ref{eqn:fairness}).\label{eqn:solveForAccess}
\end{align}
Since the power vector is given, the problem~(\ref{eqn:solveForAccess}) can be separated into maximizing throughputs $\throughputCellCh{m}{s}$ for each selection of cell $m$ and channel $s$, where $s$ is a channel in the set of channels assigned to $m$. For every such selection of $m$ and $s$, we must choose access probabilities $\accessPerChannel{s}{i}$, for every node $i$ in cell $m$, such that $\throughputCellCh{m}{s}$ is maximized. For a selection of $m$ and $s$, the maximization problem can be reduced to the following minimization.
\SG{
\footnotesize
\begin{align}
\small\text{Minimize:}&\frac{1-\accessPerChannel{s}{j}}{\accessPerChannel{s}{j}}\sigma+\left(\prod\limits_{k=1}^{ \mathcal{N}_m }\left(\frac{\RPerChannel{s}{k}{k'}}{\RPerChannel{s}{j}{j'}} +\frac{1-\accessPerChannel{s}{j}}{\accessPerChannel{s}{j}}\right)-\frac{1-\accessPerChannel{s}{j}}{\accessPerChannel{s}{j}}\right)\Tcol,\hspace{-1em}\label{eqn:access1}\\
\small\text{subject to:}&\enspace 0 \leq\accessPerChannel{s}{i}\leq 1.\label{eqn:access2}
\end{align}}
\SG{The reduction can be obtained by using the fairness constraint~(\ref{eqn:fairness}) to rewrite the throughput $\throughputCellCh{m}{s}$ of cell $m$ on channel $s$ in terms of the access probability $\accessPerChannel{s}{j}$ of node $j$ in the cell that has the smallest payload rate $\RPerChannel{s}{j}{j'}$ on $s$ amongst all nodes in the cell. The problem~(\ref{eqn:access1})-(\ref{eqn:access2}) is one of convex optimization. See Appendix~\ref{appendix:A} for details.} Solving the problem gives us the access probability $\accessPerChannel{s}{j}$ for the chosen node $j$. This probability together with the fairness constraint~(\ref{eqn:fairness}) can be used to calculate the corresponding access probabilities for all other nodes in the cell.

\subsubsection{Solve for Power}
Given a vector $\mathbf{\Psi}^*$ that solves~(\ref{eqn:solveForAccess}), we solve for a vector of transmit powers that optimizes the network throughput. The optimization problem is
\begin{align}
\text{Maximize:}\enspace T(\mathbf{\Psi}^*,\textbf{P}),\quad {\text{subject to:}}\ (\ref{eqn:maxConst}),~(\ref{eqn:sumPower}),~(\ref{eqn:fairness}).
\label{eqn:solveForPower}
\end{align}
The problem is non-convex in $\textbf{P}$. \SG{To show this, observe that the equality constraint~(\ref{eqn:fairness}) is non-linear in $\textbf{P}$.} 
\section{Evaluation Methodology}
\label{sec:simulation}
\begin{figure*}[t]
\begin{center}
  \subfloat[]{\includegraphics[scale=0.15]{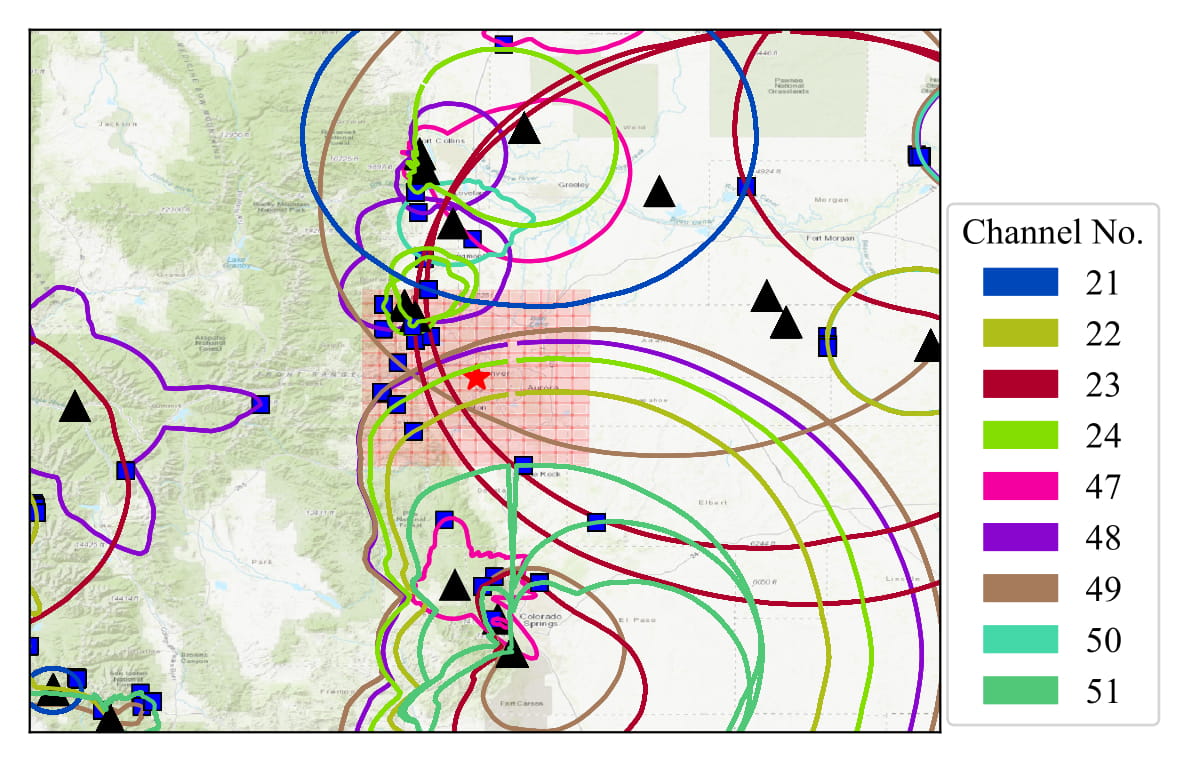}\label{fig:simulation_net_denver}}
  \enspace
  \subfloat[]{\includegraphics[scale=0.15]{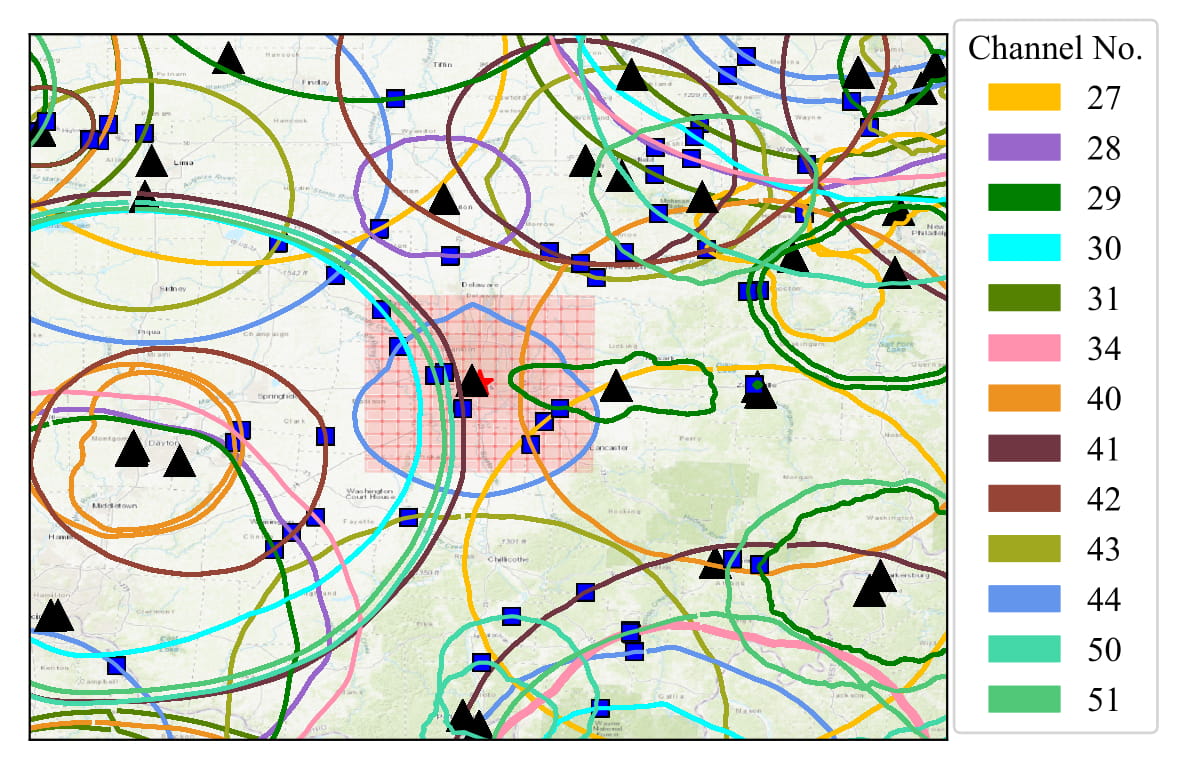}\label{fig:simulation_net_columbus}}
  \enspace
  \subfloat[]{\includegraphics[scale=0.15]{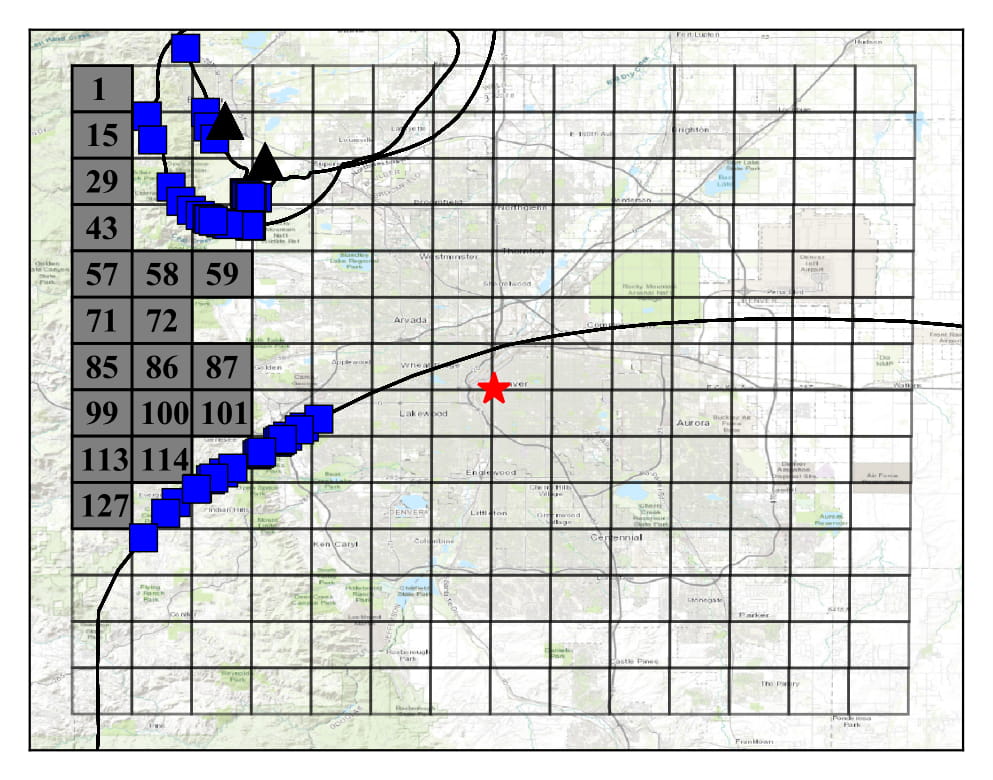}\label{fig:WorstcaseReceiver}}
  \caption{\small Illustration of White-Fi networks deployed over an area of $4900$ km$^2$ in the cities of (a) Denver and (b) Columbus. City centers are marked by a solid star. Each White-Fi network comprises of $196$ cells where each cell is spread over an area of $25$ km$^2$. TV networks operating on different available channels are also shown. Each TV network comprises of a TV tower (solid triangle) surrounded by service contour. For each tower, we show a TV receiver (solid square) located on its service contour. The diagonal of the shown maps is of length $800$ km. (c) Illustration of afflicted TV receiver (solid blue squares) locations on the service contour of TV transmitters operating on channel $24$ in the city of Denver. The solid black triangles are the TV transmitters (only two are seen in the figure). Cells colored grey are the ones at which channel $24$ is available. Each cell covers an area of $25$ km$^2$.}
  \label{fig:simulation_net}
\end{center}
\vspace{-1.5em}
\end{figure*}

We describe the White-Fi and TV networks that we used to evaluate our approach.

We considered hypothetical deployments of White-Fi networks over the cities of Denver and Columbus in the United States. Each White-Fi network was deployed over a square region of area $4900$ km$^{2}$ that covers the city. The region was tessellated by squares. Each square was considered to be a separate White-Fi cell. For the sake of evaluation, we considered networks tessellated by cells of areas $12.25$ (square of length $3.5$ km), $25$, and $100$ km$^{2}$. These areas correspond to, respectively, $400$, $196$, and $49$ White-Fi cells in the White-Fi network. White-Fi networks tessellated by cells of area $25$ km$^2$ and superimposed on the maps of the cities are shown in Figure~\ref{fig:simulation_net}. For a given channel availability and White-Fi node power budget, one would expect throughput to deteriorate as the cell size increases. In practice, however, large cell sizes may be desirable in sparsely populated areas with limited access to wired backhaul connectivity to the Internet.

We simulated a total of $4900$ White-Fi nodes that were split equally amongst all cells. Nodes in each cell were distributed uniformly and independently of other nodes. For every node in a cell, another node in the cell was chosen randomly as the receiver of its data payload. Each node was assigned a total power budget $P_T = 0.1$ W. This is also the maximum transmit power that FCC allows personal/portable nodes operating outside the protected region of a TV transmitter. As we will show later, nodes in cells close to TV networks operating on their assigned channels are often unable to exhaust this assigned power budget. For the chosen cell sizes and power budget, cells are adjacent if and only if they are physically adjacent (have a common edge). 

All cells use a bandwidth of $6$ MHz centered around each assigned channel. For this bandwidth, the $802.11$ timing parameters were obtained by scaling, by a factor of about $3$, the parameters used in~\cite{bianchi} for a WiFi network that uses $20$ MHz of bandwidth. The length of payload was set to $L=8184$ bits. All link gains were calculated using a \SG{simplified} path loss model~\cite{goldsmith} with exponent $3$.

We obtained information about the TV networks in and around the region covered by the White-Fi networks from~\cite{fccdata},~\cite{fccservicecontourpoints}, and~\cite{spectrumbridge}. Information such as TV transmitter locations, their operating channel, and transmit powers, was obtained from the FCC database~\cite{fccdata}. Service contours were obtained from~\cite{fccservicecontourpoints}. We used them to create the protected region for every TV TX by following the guidelines in~\cite{spectrumbridge}. We have compiled a database~\cite{sneihil2017} that includes all the above information for all TV transmitters in the US. Figures~\ref{fig:simulation_net_denver} and~\ref{fig:simulation_net_columbus} show the resulting TV transmitters, their service contours, and the channels in which they operate, for the cities of Denver and Columbus, respectively.  

We considered only TV channels in the set $\{21,\dots,51\}\setminus\{37\}$ for assignment to White-Fi nodes. This is as per the FCC regulations for personal/portable devices. These channels occupy the spectrum in the $512-698$ MHz range. We evaluated our approach for the following two ways of calculating channels available in a White-Fi cell.

\begin{enumerate}
\item \emph{Exact} FCC: A channel may be accessed by a node only if it lies outside the protection region of all TV transmitters that operate on the channel.
\item \emph{Relaxed}: A node may be inside the protection region. It must, however, be outside the service contour of all TV transmitters on the channel.
\end{enumerate}

While location information is available for TV transmitters, it is not available for TV receivers. As a workaround, for each TV transmitter and White-Fi cell assigned the channel on which the transmitter broadcasts, we placed a TV receiver at a location deemed to be most afflicted by interference from nodes in the cell. At such a location, the constraint~(\ref{eqn:maxConst}) on aggregate interference is binding, for a fixed limit $\Imaxlimit{l}$ for all receiver locations $l$.

Given the \SG{simplified} path loss model, this \emph{most afflicted} receiver must lie on the service contour of the TV transmitter. Further, we approximated the most afflicted location by calculating for each vertex of the cell the point on the contour that is closest to it. Among the four obtained points, we picked the point that has the smallest distance from its corresponding vertex as the location of the \emph{most afflicted} receiver. The resulting locations of receivers operating on channel $24$ in the city of Denver are shown in Figure~\ref{fig:WorstcaseReceiver}.

In all evaluation, we assumed a limit on maximum interference $\Imaxlimit{l}= -140\text{ dB}, \forall l$. This was obtained by assuming that the TV receiver is tolerant to about $3$ dB increase in its noise floor~\cite{sahai}. \SG{While there are works on WiFi networks in TV white spaces~(\cite{radunovic,KTH13} and~\cite{wifiXL}), they do not capture the CSMA/CA based mechanism of the DCF, which is key to our problem. For the purpose of baselining,} in the following section, we compare our \emph{proposed} approach detailed in Section~\ref{sec:method} with a \emph{baseline} that doesn't leverage the heterogeneity in available link quality, due to the presence of the TV network, within a cell. Specifically, the \emph{baseline} allocates the same power and access probabilities to all nodes in a cell on an assigned channel. This allocation may, however, vary over channels assigned to the cell and over different cells. Both \emph{baseline} and \emph{proposed} use the same channel assignment.
\section{Results}
\label{sec:results}
We use our proposed approach to compare the White-Fi throughput obtained in the cities of Denver and Columbus. We show how greater channel availability in Columbus than in Denver doesn't translate into larger throughput in Columbus due to a much larger presence of TV transmitters in the city. We also show the throughput achieved by the \emph{baseline} that makes power and access probability allocation within a cell homogeneous and hence a lot simpler in practice. We end this section with an estimate of loss of White-Fi coverage when White-Fi access is allowed by rules akin to FCC regulations, wherein \emph{all} nodes outside a certain region around a TV transmitter are allowed to transmit using their full power budget of $100$ mW.
\subsection{Observations on Network Throughput}
\begin{figure}
\vspace{-9em}
\begin{center}
\includegraphics[width=0.7\linewidth,bb = 50 50 1000 1000]{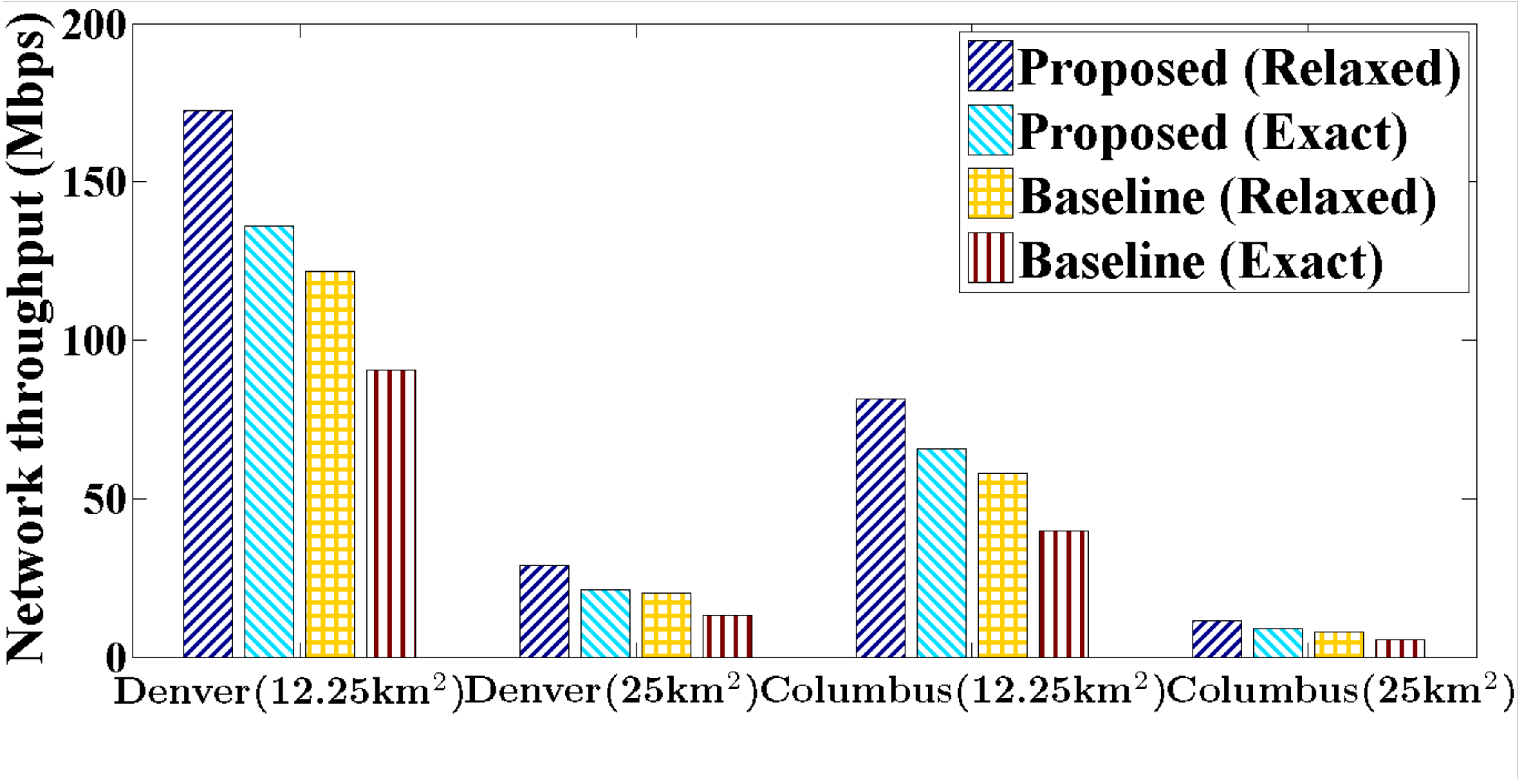}
\end{center}
\caption{\small Throughputs obtained by the \emph{proposed} method and the \emph{baseline} for the hypothetical White-Fi deployments in Denver and Columbus shown in Figure~\ref{fig:simulation_net}. We show throughputs for cell size choices of $12.25$ and $25$ km$^2$, and channel availability calculated using \emph{Exact} FCC and \emph{Relaxed}. The corresponding throughputs (kbps) for a cell size of $100$ km$^2$ are $815$, $644$, $530$, and $362$ for Denver and $260$, $197$, $197$, and $124$ for Columbus.}
\label{fig:NetThr}
\vspace{-1em}
\end{figure}

\emph{As a Function of Cell Size:} Figure~\ref{fig:NetThr} shows the network throughputs~(\ref{eqn:netwkThr}), obtained using the proposed approach and the baseline, for the cities of Denver and Columbus. For both approaches and cities, the throughput reduces with increasing cell size. This is because in larger cells, White-Fi nodes are typically farther apart and see smaller link gains than in smaller cells. While the median link gains for a cell size of $12.25$ km$^2$ are about $-90$ dB, they are $10$ dB less for a cell size of $100$ km$^2$. Note that, as we will see later, channel assignment is not impacted by choice of cell size.
\emph{Baseline vs. Proposed:} In both the cities, for different cell sizes, and the approaches of \emph{Exact} FCC and \emph{Relaxed}, the proposed approach leads to large throughput gains in the range of $40-70\%$ over baseline. 

\emph{Exact \emph{FCC} vs. Relaxed:} Now compare the throughputs obtained when using \emph{Exact} FCC and \emph{Relaxed}. The choice of \emph{Relaxed} always leads to larger throughputs. For a cell size of $12.25$ km$^2$ and the city of Denver, \emph{Relaxed} leads to gains in throughput of about $27\%$ over the throughput obtained using \emph{Exact} FCC. For Columbus, the corresponding gains are about $36\%$. Similar gains in throughput are seen for larger cell sizes too. In fact, even the \emph{Baseline} approach leads to similar gains on using \emph{Relaxed}. Recall that \emph{Relaxed} allows White-Fi nodes to utilize a channel as long as the nodes are outside the service contour of the TV transmitter using the channel. \emph{Exact} requires nodes to be outside the protection region. That is the channel cannot be used over a larger region. This impacts channel availability and, hence, channel assignment to White-Fi cells, and explains the observed reduction in network throughput.

\begin{figure*}
\vspace{-5em}
  \centering
  \subfloat[]{\includegraphics[width=0.35\linewidth,bb = 0 0 900 700]{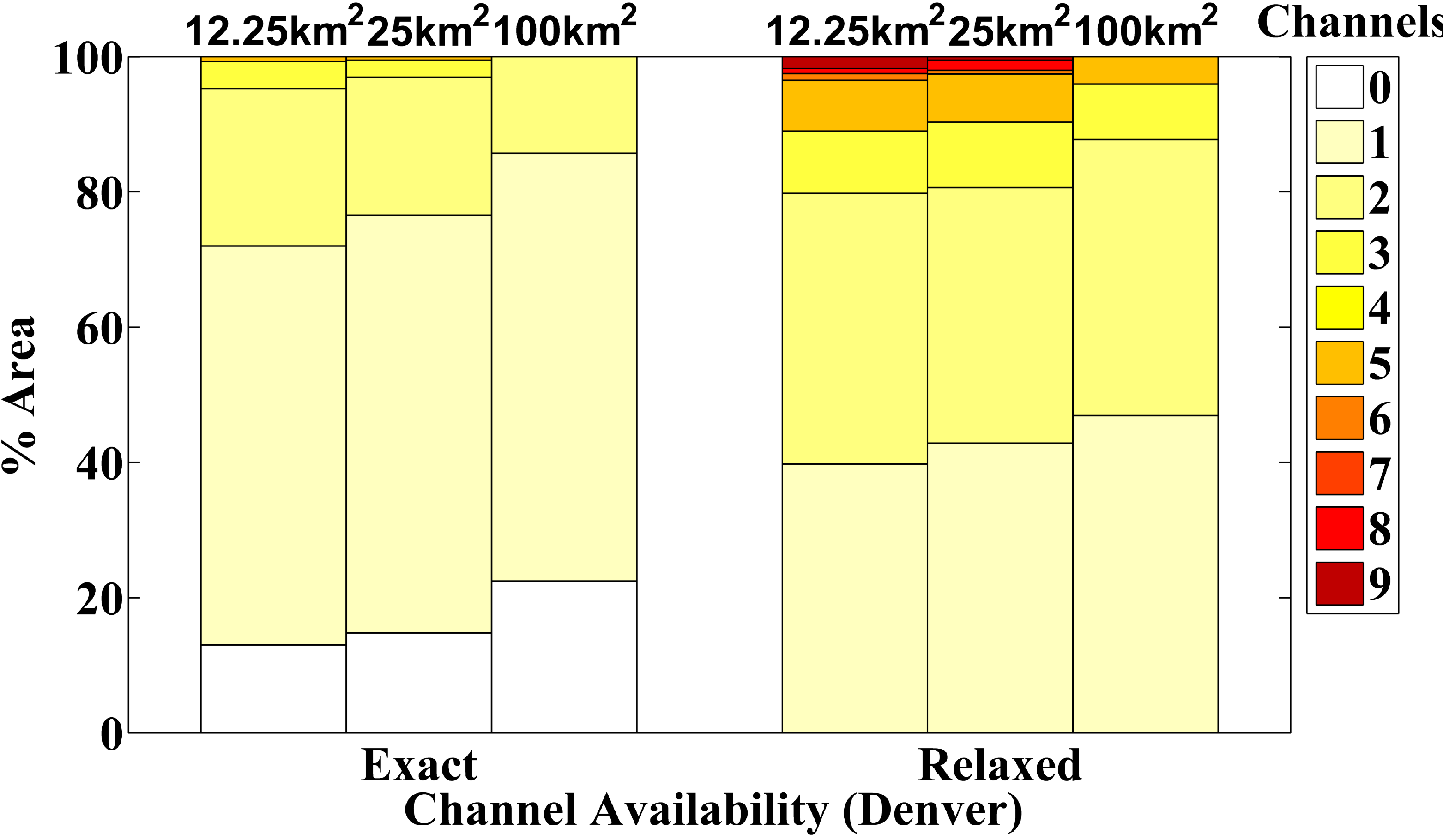}\label{fig:PercAreaChanAvlDenver}}
  \quad
  \subfloat[]{\includegraphics[width=0.35\linewidth,bb = 0 0 900 700]{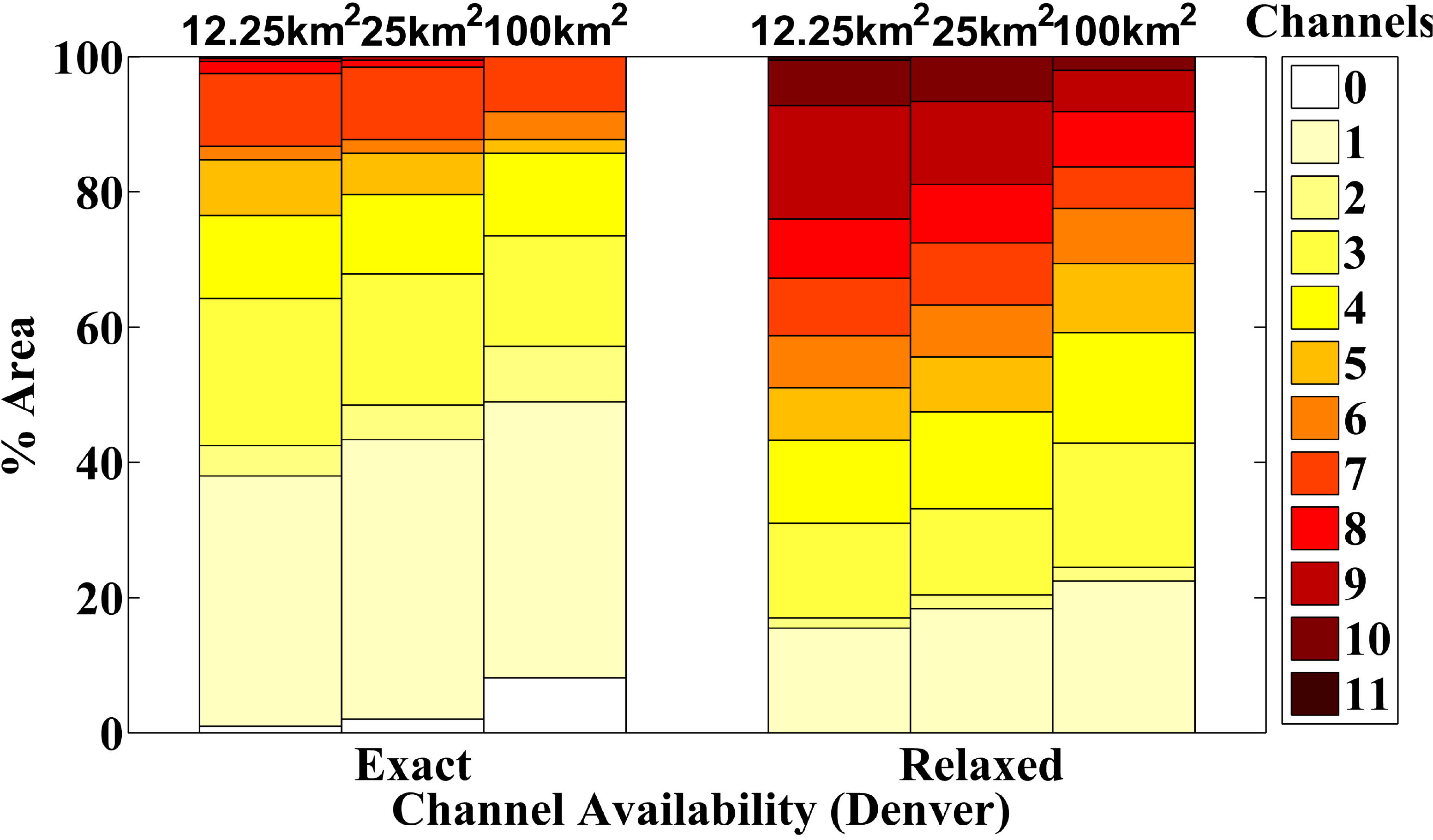}\label{fig:PercAreaChanAvlColumbus}}\\
  \vspace{-5em}
    \subfloat[]{\includegraphics[width=0.35\linewidth,bb = 0 0 900 700]{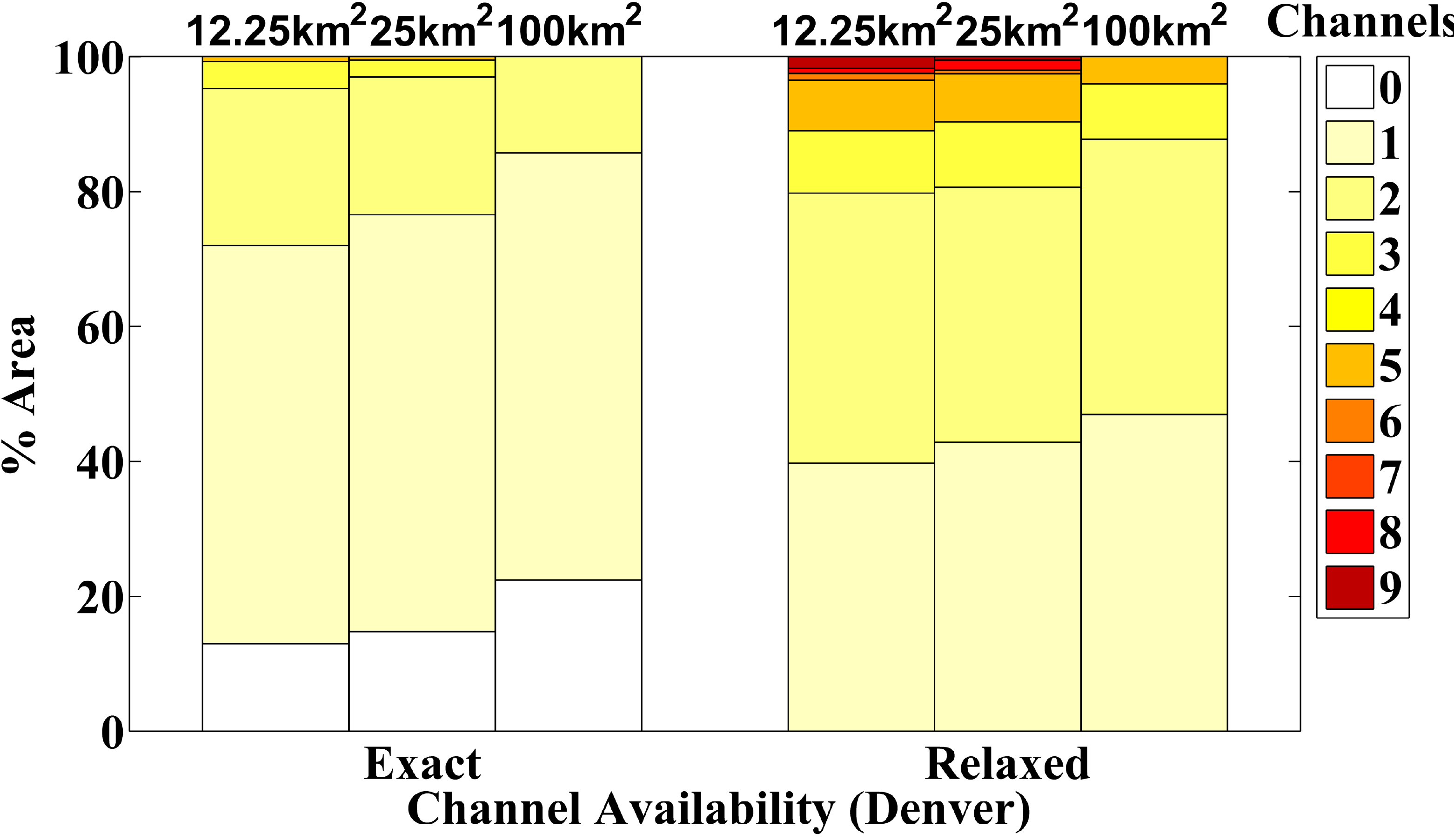}\label{fig:PercAreaChanAssignDenver}}
  \quad
  \subfloat[]{\includegraphics[width=0.35\linewidth,bb = 0 0 900 700]{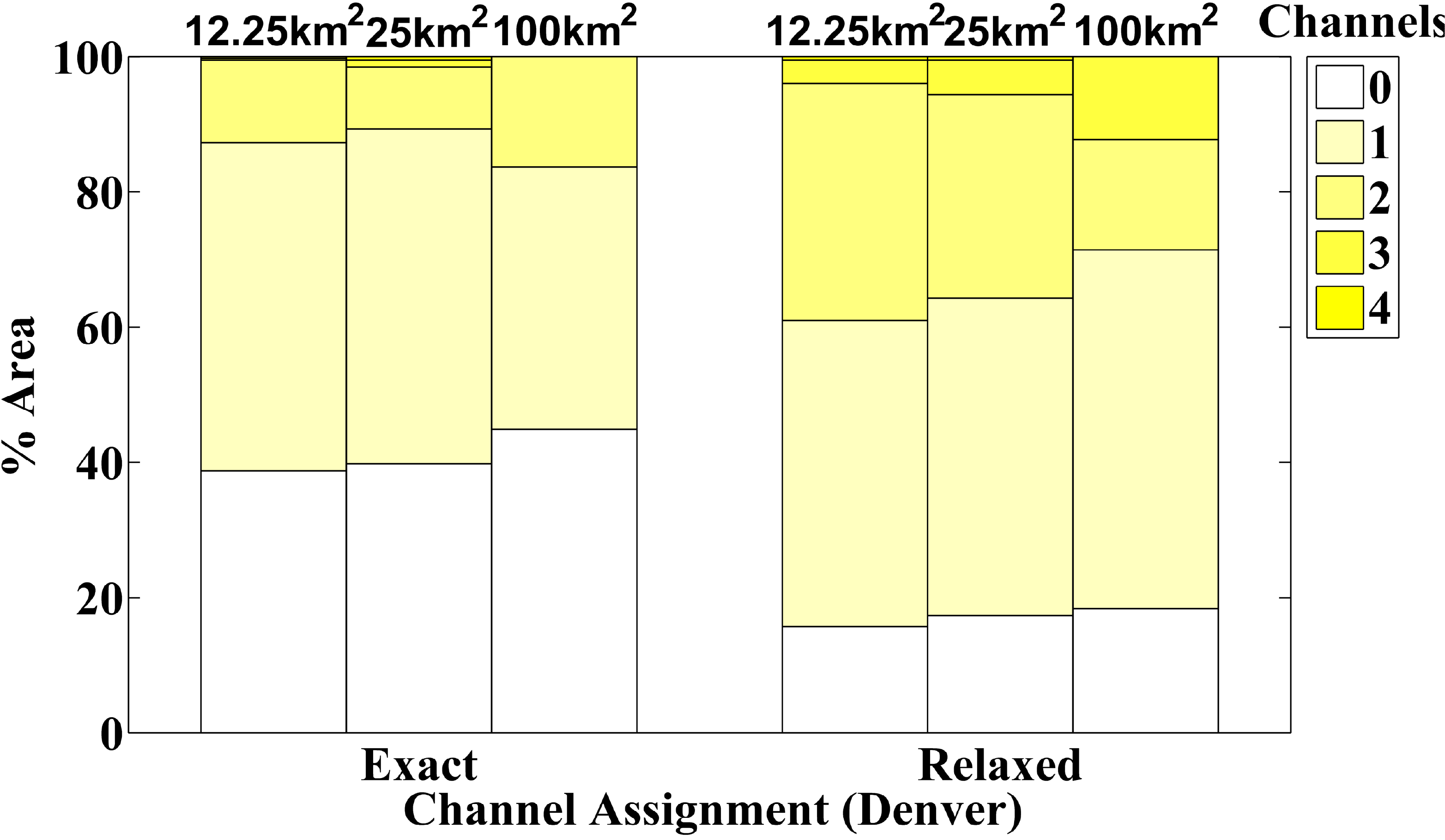}\label{fig:PercAreaChanAssignColumbus}}
\caption{\small Distribution of the number of available ((a) and (b)) and assigned ((c) and (d)) channels over the White-Fi networks in Denver and Columbus.}
\label{fig:PercArea}
\vspace{-1em}
\end{figure*}
\emph{Channel Availability and Assignment:} Figures~\ref{fig:PercAreaChanAvlDenver} and~\ref{fig:PercAreaChanAvlColumbus} show for the cities of Denver and Columbus, respectively, the distribution of the number of available channels over area covered by the White-Fi network. For example, for a cell size of $12.25$ km$^2$ and the city of Columbus, when using \emph{Relaxed}, about $20\%$ of the area has exactly one available channel and about $60\%$ has greater than $4$ available channels. Both Denver and Columbus have a larger average number of channels available per unit area when using \emph{Relaxed}. Specifically, when using \emph{Relaxed}, on an average Columbus has $5$ available channels as opposed to $3$ when using \emph{Exact} FCC. The corresponding numbers for Denver are $2$ and $1$. This results in a larger number of \emph{assigned} channels, when using \emph{Relaxed}, as shown in Figures~\ref{fig:PercAreaChanAssignDenver} and~\ref{fig:PercAreaChanAssignColumbus}.

In Denver, more than $50\%$ of the area under the White-Fi network, for both \emph{Relaxed} and \emph{Exact} FCC, is not assigned any channel. Most cells in the remaining area are assigned a single channel. This is explained by the fact that most cells in the network have either channel $21$ or $51$ available and the channel assignment constraint~(\ref{eqn:adjacency}) must be satisfied. In Columbus, a smaller region suffers from outage. Especially under \emph{Relaxed}, greater than $80\%$ of the region is assigned at least one channel and about $40\%$ (for a cell size of $12.25$ km$^2$) is assigned two or more channels. Again, the numbers of assigned channels are much smaller than the corresponding numbers of available channels. Given the assignment constraint~(\ref{eqn:adjacency}), this is explained by the fact that the total number of unique available channels is just $12$ and, as is seen in Figure~\ref{fig:channelaval}, cells with large numbers of available channels are clustered together in space.

\emph{Network Throughput of Columbus is Smaller than that of Denver:} As observed above, when compared to Denver, a much smaller area of Columbus is starved of white space channels and a larger percentage of area is assigned more than one channel. However, Columbus has a network throughput (see Figure~\ref{fig:NetThr}) much smaller than Denver. For example, for a cell size of $12.25$ km$^2$ and using \emph{Relaxed}, the throughput of Columbus, using \emph{Proposed}, is about half that of Denver.

\begin{figure*}
\vspace{-6em}
  \centering
  \subfloat[]{\includegraphics[width=0.31\linewidth,bb = 0 0 900 900]{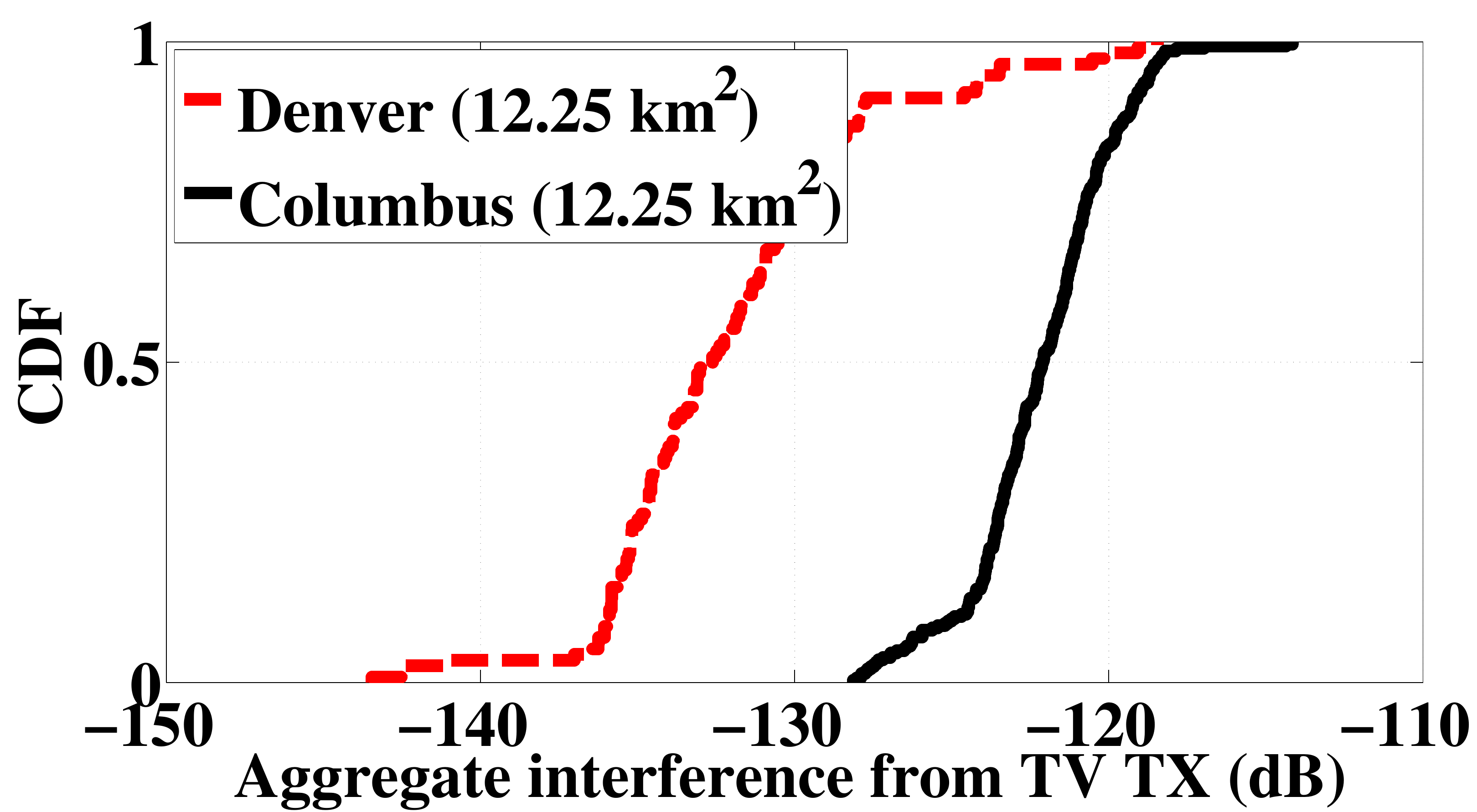}\label{fig:CDFIntfromTV}}
  \quad
  \subfloat[]{\includegraphics[width=0.31\linewidth,bb = 0 0 900 900]{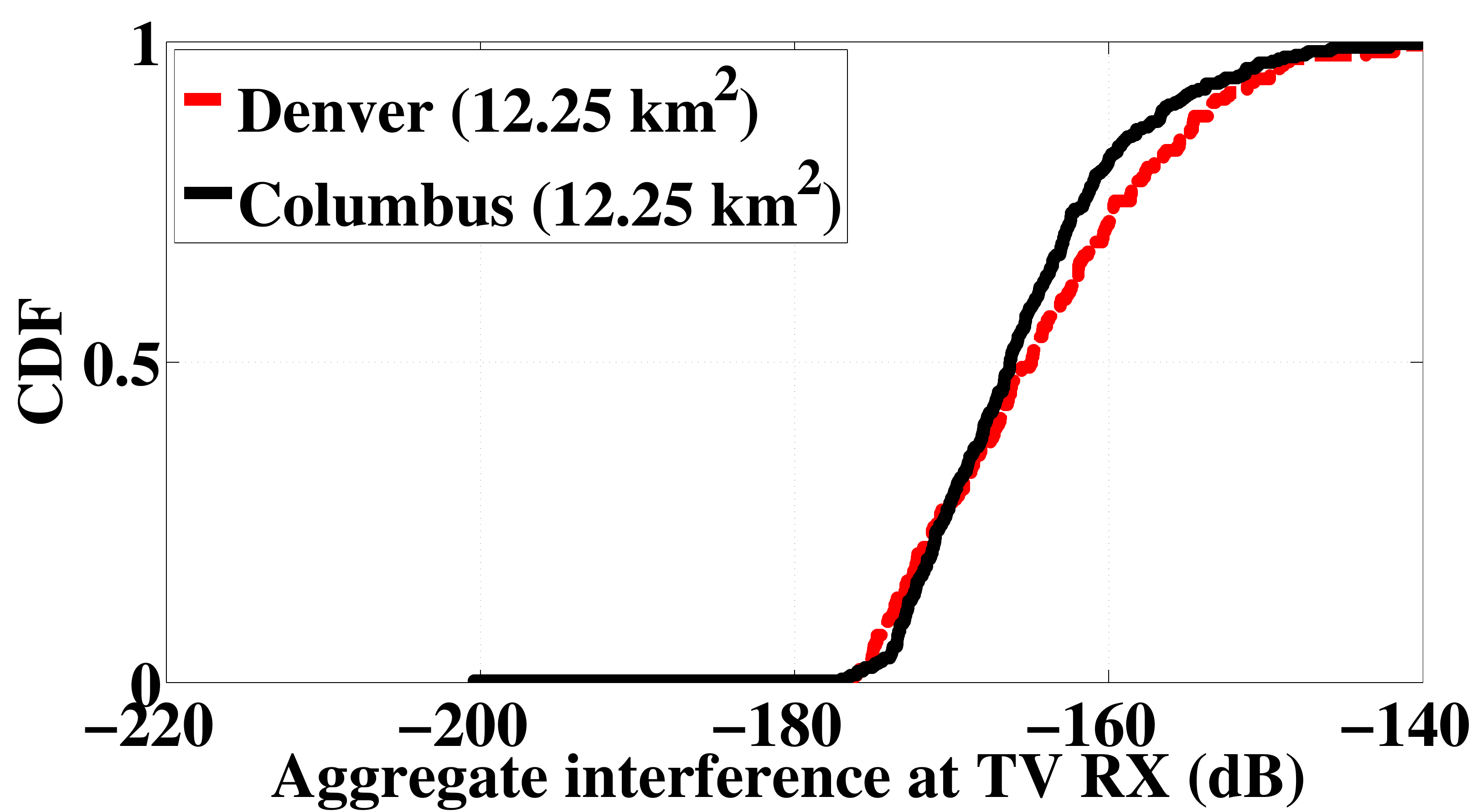}\label{fig:CDFInttoTV}}
  \quad
  \subfloat[]{\includegraphics[width=0.31\linewidth,bb = 0 0 900 900]{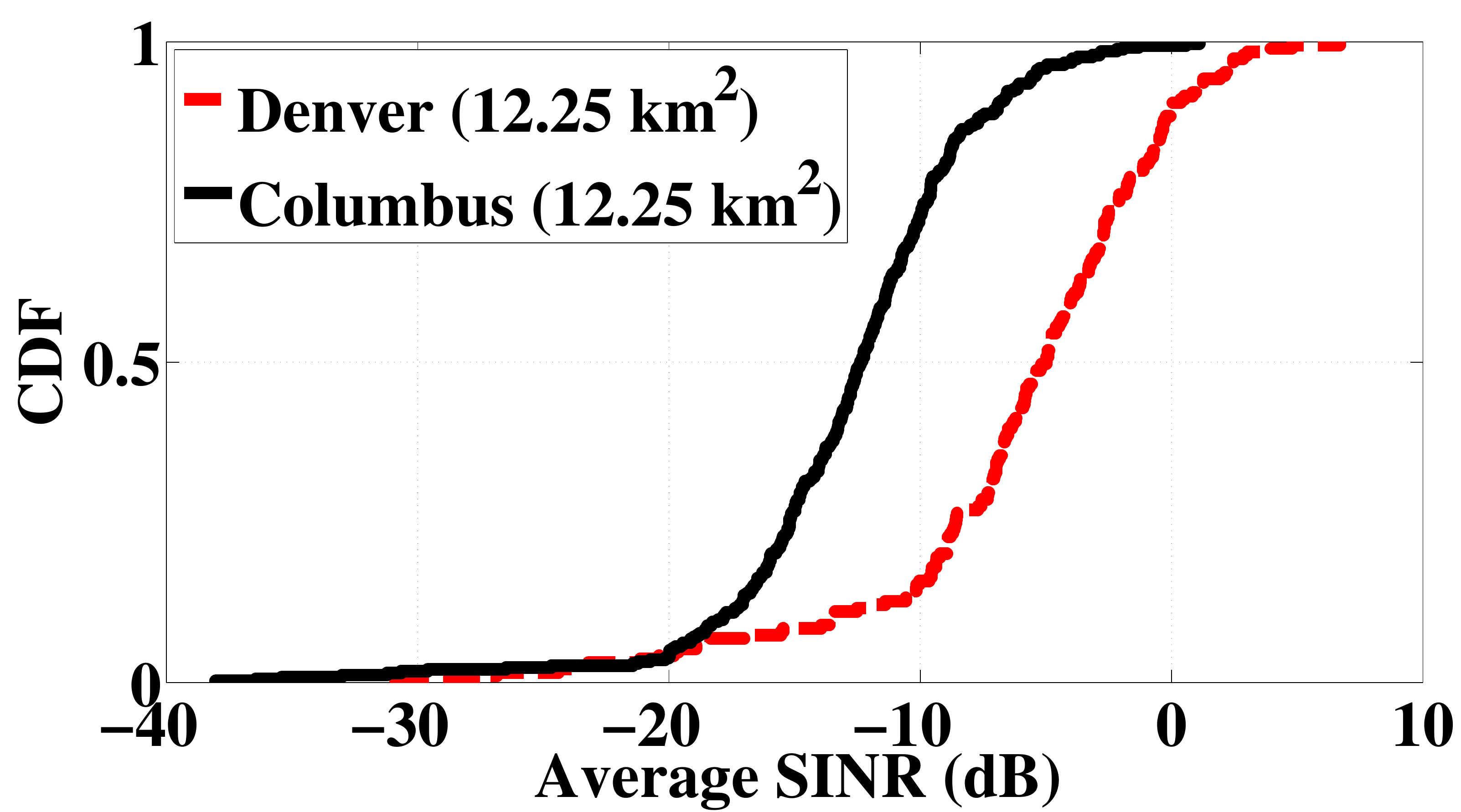}\label{fig:CDFSINR}}
  \caption{\small Empirical CDFs of (a) average aggregate interference from TV transmitters at White-Fi nodes in a cell (b) aggregate interference at afflicted TV receivers from White-Fi nodes in a cell (c) average SINR(dB) of nodes in a cell.}
\label{fig:IntfromTV}
\vspace{-1em}
\end{figure*}

It turns out that while Columbus has more assigned channels on an average, it also suffers significantly more due to a high density of TV networks (see Figures~\ref{fig:simulation_net_denver} and~\ref{fig:simulation_net_columbus}). Figures~\ref{fig:CDFIntfromTV}-\ref{fig:CDFSINR} compare the impact of the TV networks on the White-Fi networks in Denver and Columbus.
Figure~\ref{fig:CDFIntfromTV} shows the empirical CDF (cumulative distribution function) of aggregate interference from TV transmitters averaged over White-Fi nodes in a cell, for each of the two cities. For a cell that is assigned more than one channel, the aggregate is chosen for an assigned channel on which it is the minimum. Nodes in Columbus see aggregate interference that, on an average, is about $10$ dB larger than that seen by nodes in Denver.

Figure~\ref{fig:CDFInttoTV} shows the CDF of the aggregate interference created by White-Fi nodes at afflicted TV receivers operating on the assigned channels that were selected for Figure~\ref{fig:CDFIntfromTV}. The CDF(s) for both the cities are similar, which says that the TV receivers are as much of a constraint in Denver as in Columbus. This makes us believe that the reduced throughput seen by Columbus is because of excessive interference from TV transmitters. The impact of the TV transmitters and receivers is summarized in Figure~\ref{fig:CDFSINR} that shows the CDF of SINR of White-Fi links, for each of the two cities. Links in Columbus see SINR that is on an average about $7$ dB smaller than SINR of links in Denver.

We end our observations on throughput by noting that channel availability reduces slightly with increasing cell size in Figures~\ref{fig:PercAreaChanAvlDenver} and~\ref{fig:PercAreaChanAvlColumbus}. This is because a channel is said to be available in a cell only if it is available in all of the area of the cell. If assigned, such a channel may be used by any node in the cell. Our method of calculating availability, however, has little or no impact on assignment. In fact, the reduction in throughout with cell size is, as explained above, simply a result of smaller link gains in larger cell sizes.
\subsection{Explaining Gains on Using the Proposed Approach}
\begin{figure*}
\sbox\twosubbox{%
  \resizebox{\dimexpr 0.95\textwidth-1em}{!}{%
    \includegraphics[height=0.2cm,scale=1,bb = 0 0 100 100]{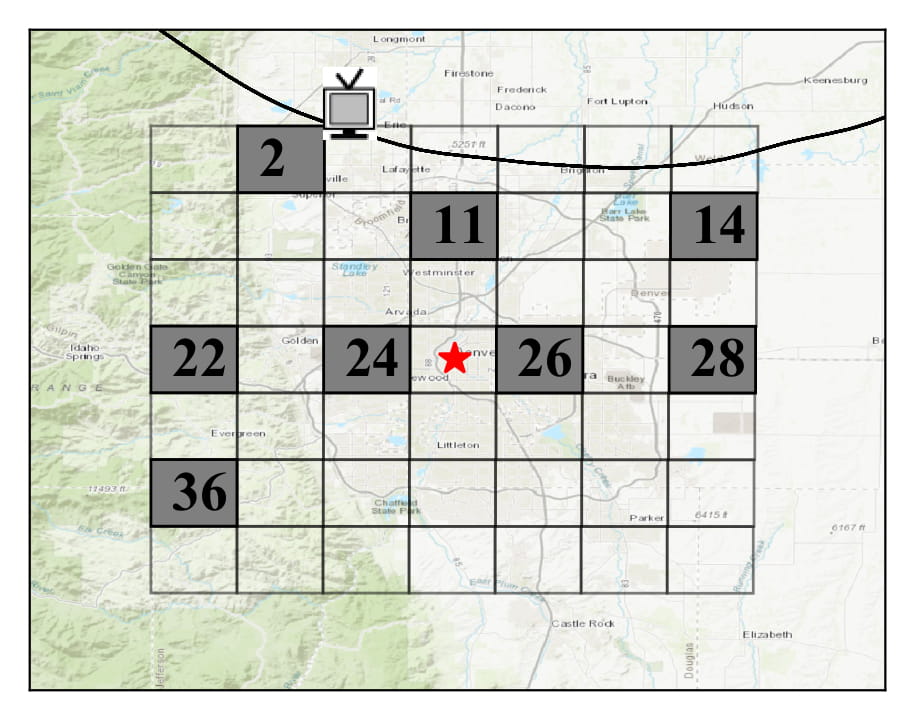}%
    \includegraphics[height=0.2cm,scale=1,bb = 0 0 100 100]{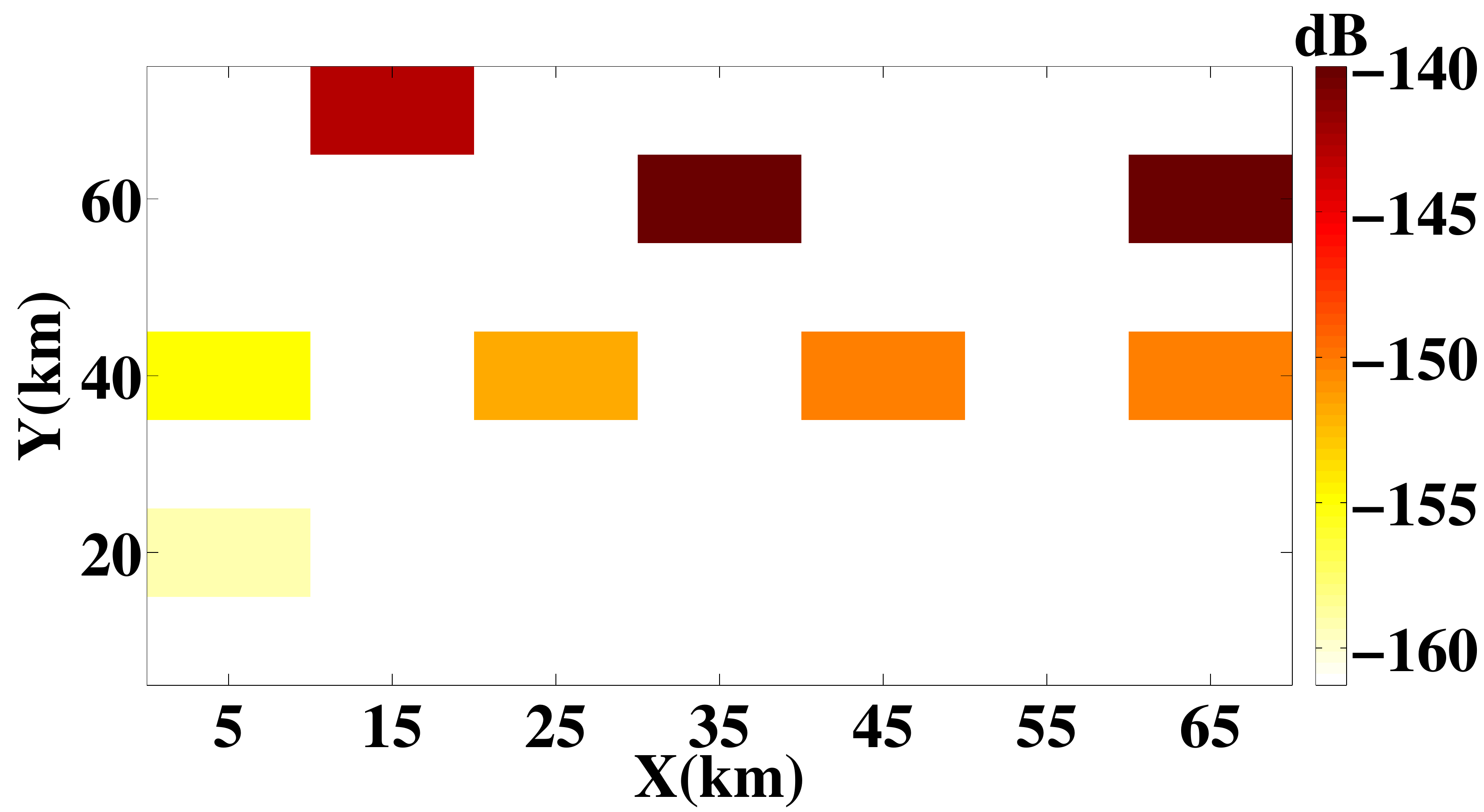}%
    \includegraphics[height=0.2cm,scale=1,bb = 0 0 100 100]{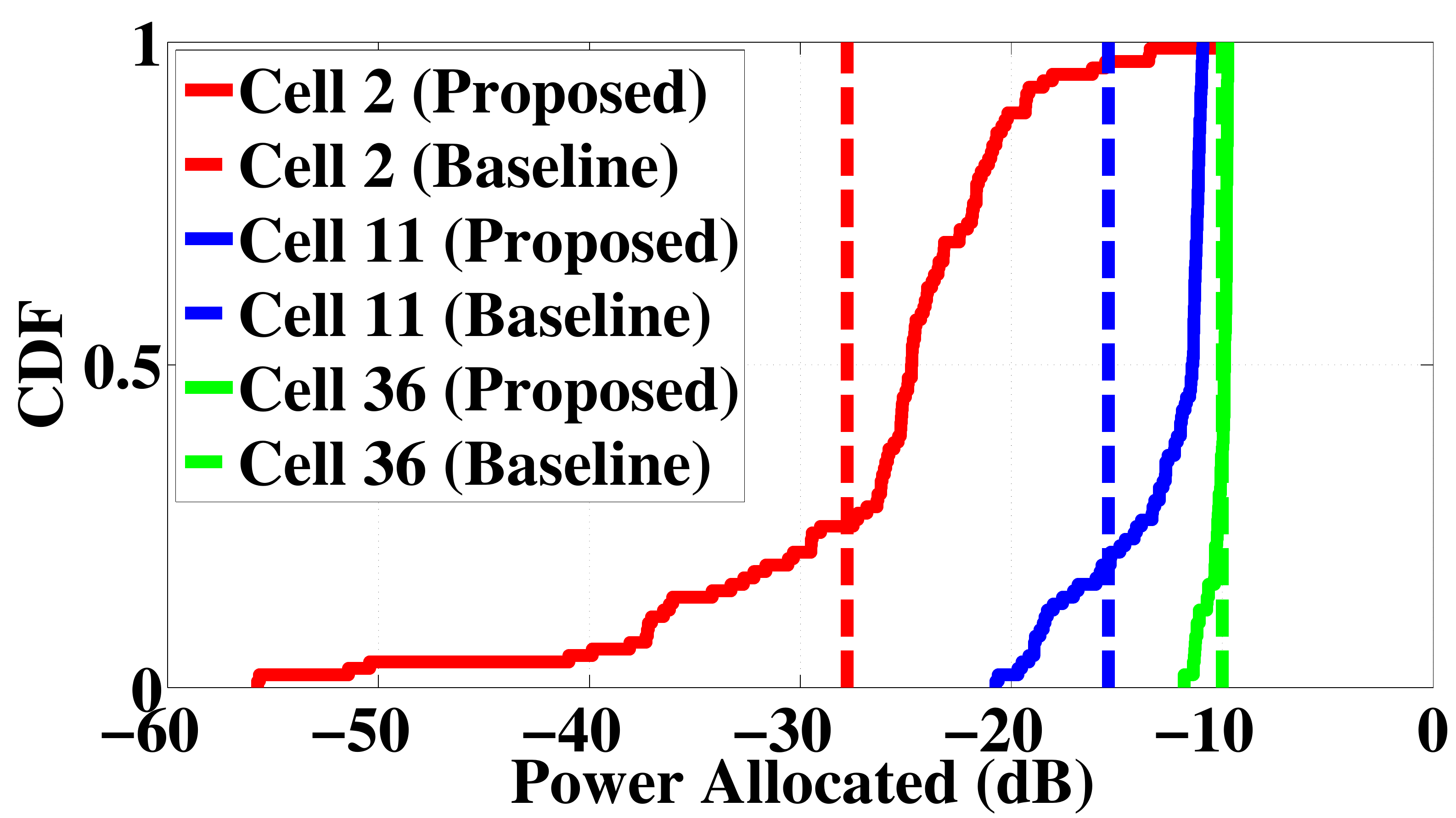}
    \vspace{2cm}
  }%
}
\setlength{\twosubht}{\ht\twosubbox}
  \centering
  \captionsetup[subfloat]{margin=-1.8cm}
  \hspace{.8cm}
  \subfloat[]{\label{fig:WorstReceivers}}{%
    \fbox{\includegraphics[height=.6\twosubht]{Figures/TV_tower_worst_receiver_chan21_Denver.jpg}}%
  } \hspace{0.85cm}
  \captionsetup[subfloat]{margin=-2.6cm}
  \subfloat[]{\label{fig:IntBudgetSharing}}{%
    \fbox{\includegraphics[height=.6\twosubht]{Figures/Interference_budget_sharing.pdf}}%
  }\quad
  \subfloat[]{\label{fig:CDFPower}}{%
    \fbox{\includegraphics[height=.6\twosubht]{Figures/CDF_power_allocation.pdf}}%
  }
    \subfloat[]{\label{fig:HetToTV}}{%
    \fbox{\includegraphics[height=.6\twosubht]{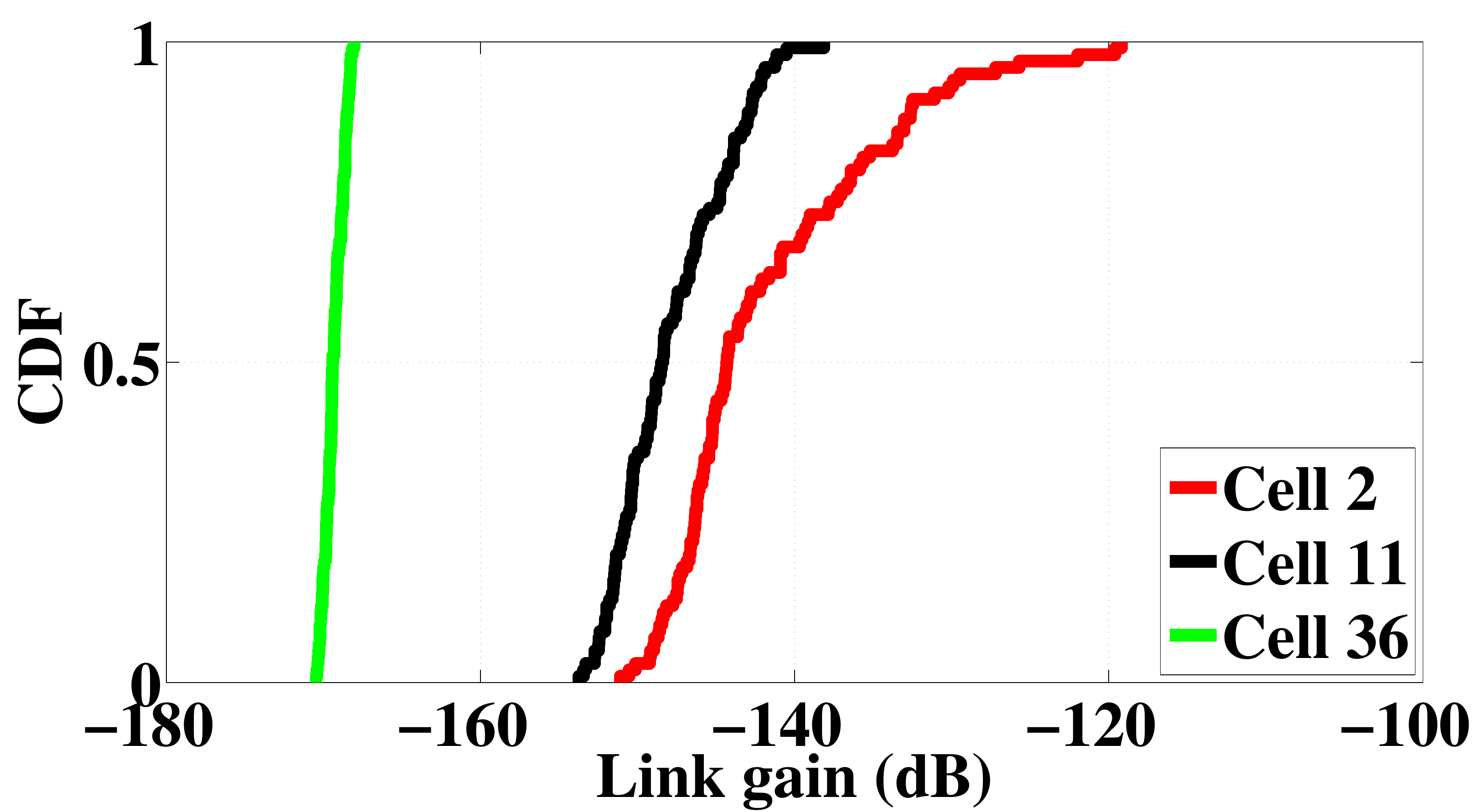}}%
  }\quad
  \subfloat[]{\label{fig:HetFromTV}}{%
    \fbox{\includegraphics[height=.6\twosubht]{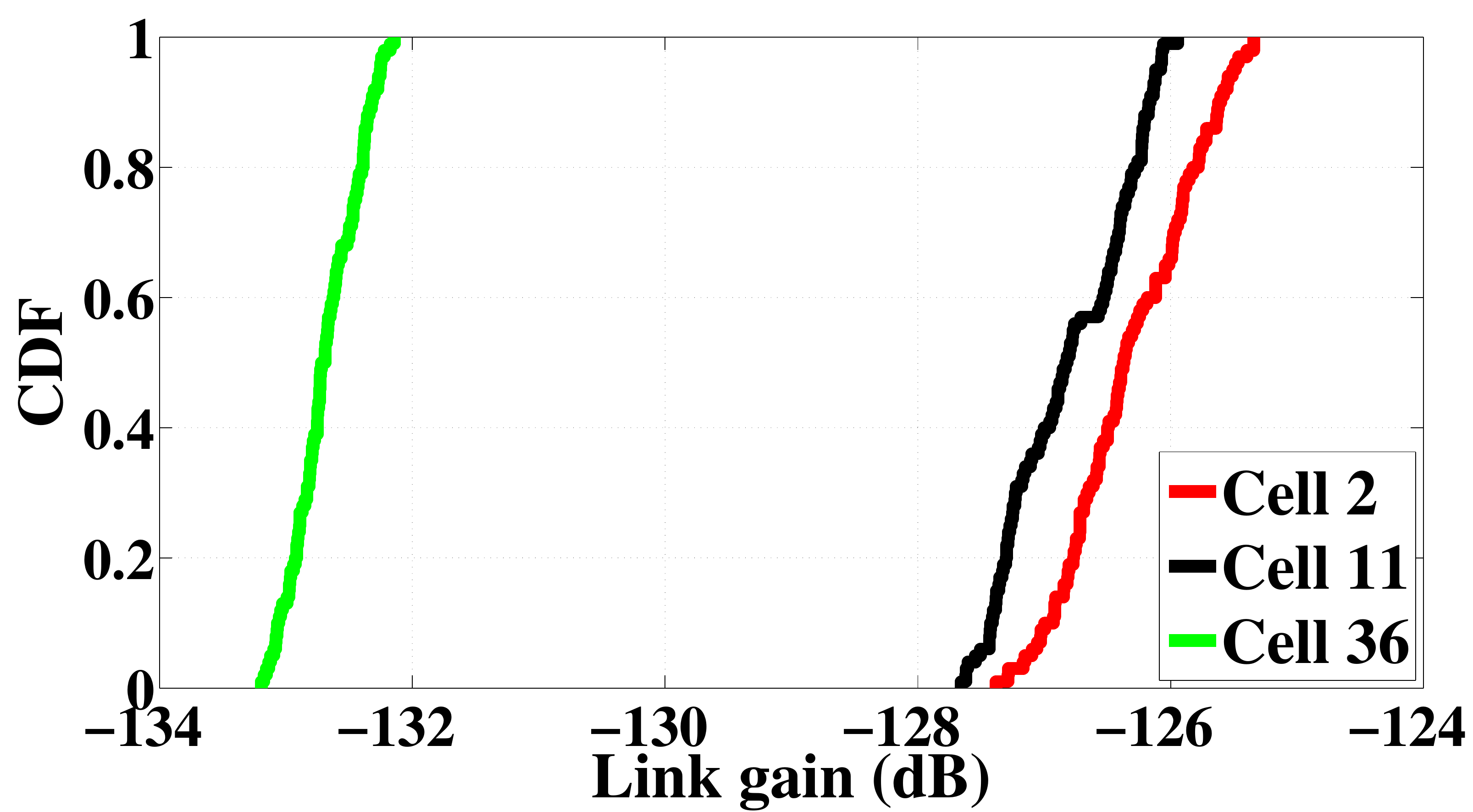}}%
  }\quad
  \subfloat[]{\label{fig:HetWithinCell}}{%
    \fbox{\includegraphics[height=.6\twosubht]{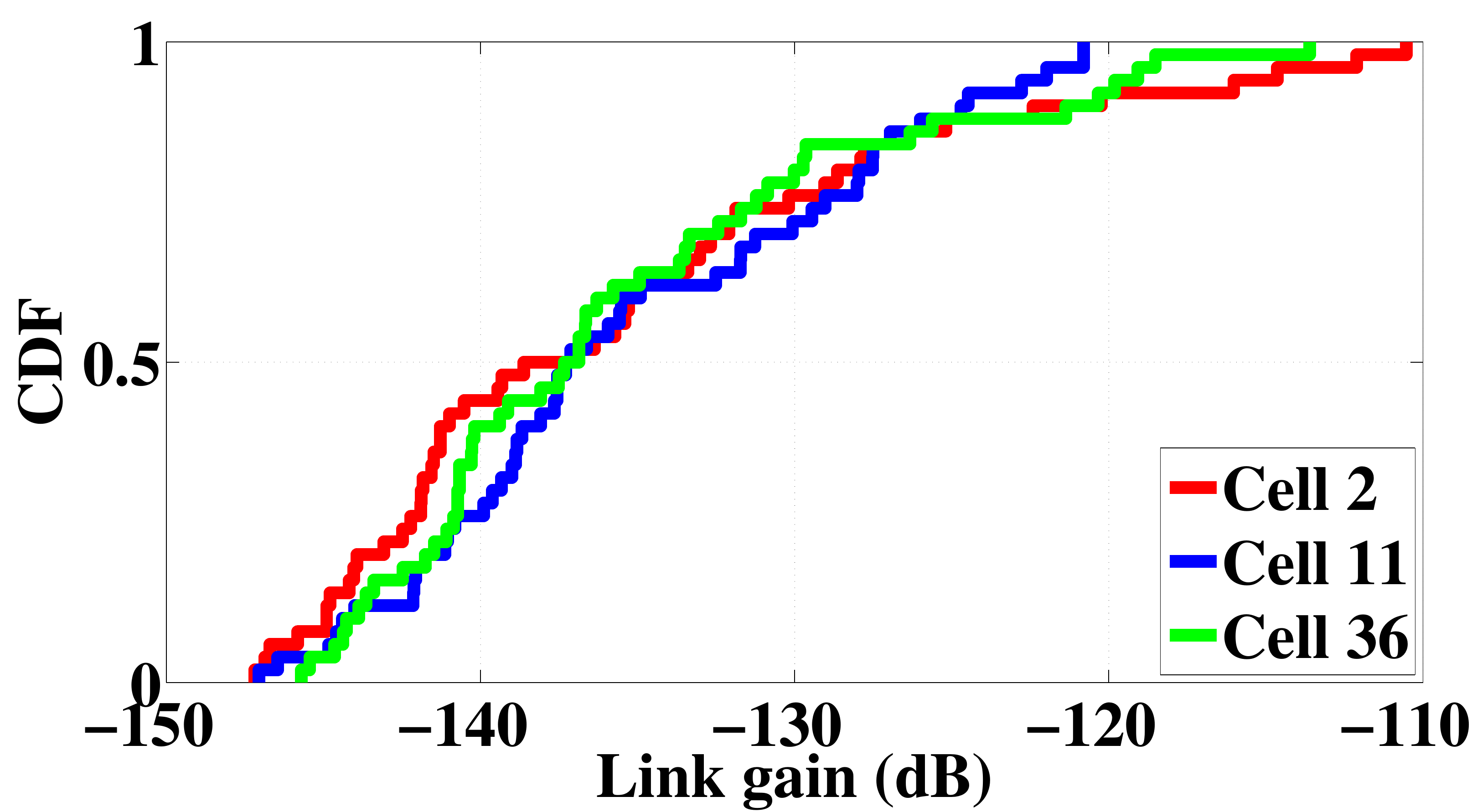}}%
  }
\caption{\small(a) We show the cells (numbered) that are assigned channel $21$ in Denver. Each cell covers an area of $100$km$^{2}$. An afflicted TV receiver is shown adjacent to cell $2$. (b) Aggregate sum interference created at the TV receiver by each of the numbered cells. (c) CDF of the power allocated to White-Fi nodes for three cells. CDF(s) are shown for the proposed method and the baseline. (d) CDF of link gains between the White-Fi nodes and the TV receiver for the three cells. (e) CDF of link gains between the White-Fi nodes and the TV transmitter broadcasting on channel $21$. The transmitter's service contour (black curve) is partly shown in (a). It is also the blue contour in Figure~\ref{fig:simulation_net_denver}. (f) CDF(s) of the link gains between White-Fi nodes for the three cells.}
\label{fig:zoomInLevelCells}  
\end{figure*}
\begin{figure}[t]
\vspace{-9em}
\begin{center}
\includegraphics[width=0.7\linewidth,bb = 100 0 950 950]{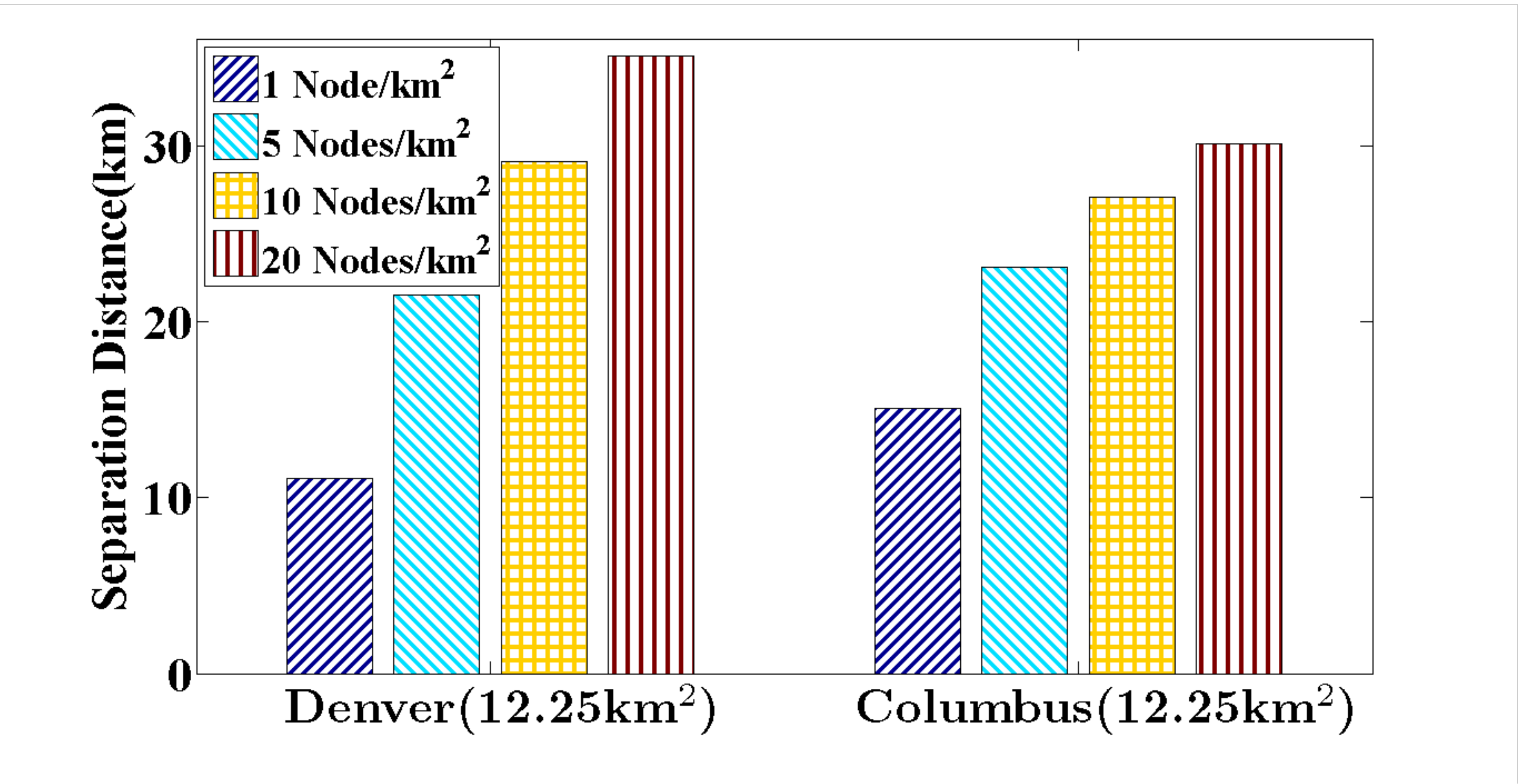}
\end{center}
\vspace{-1em}
\caption{\small Minimum desired separation distance from the service contour for varying node densities and a cell size of $12.25$ km$^2$. The distance for each density is an average calculated over multiple White-Fi node placements generated for the density.}
\label{fig:SepDis}
\vspace{-1.5em}
\end{figure}

We show how the proposed approach adapts to heterogeneity in link quality because of the TV network. We do so using cells that are assigned channel $21$ in Denver by Algorithm~\ref{alg:channelassign}. The cells are of size $100$ km$^2$. Our observations remain the same qualitatively, over other cell sizes and also Columbus. 

Figure~\ref{fig:IntBudgetSharing} shows the aggregate interference, from each of the cells assigned channel $21$, at the afflicted TV receiver in Figure~\ref{fig:WorstReceivers}. Cells that are closer to the TV receiver are responsible for larger aggregate interference and exhaust a significant share of the interference budget that is available at the TV receiver without violation of constraint~(\ref{eqn:maxConst}). To exemplify, the aggregate interference seen from cell $36$ at the TV receiver is about $-159$ dB, while the interference from nodes in cell $2$ is an aggregate of $-143$ dB.

Consider the cells numbered $2$, $11$, and $36$, in Figure~\ref{fig:WorstReceivers}. Cell $2$ is very close to the afflicted TV receiver and the service contour of the corresponding transmitter, cell $11$ is farther than $2$, and cell $36$ is the farthest. As shown in Figure~\ref{fig:HetToTV}, this results in larger link gains between the TV receiver and the White-Fi nodes in cell $2$, than for nodes in cells $11$ and $36$. Proximity of nodes in $2$ causes them to see a larger spread (greater heterogeneity) of these link gains. 

Figure~\ref{fig:HetFromTV} shows the distribution of link gains between the White-Fi nodes and the TV transmitter. Given the large service region of the TV transmitter (radius of $70.82$ km), these link gains show a limited spread of about $2$ dB for each of the three cells. While the heterogeneity in gains within a cell is limited, the three cells see different gains from the transmitter. Specifically, cell $36$ sees gains on an average about $6$ dB smaller than cells $2$ and $11$. Finally, as shown in Figure~\ref{fig:HetWithinCell}, the link gains between White-Fi nodes in the cells, as one would expect, are similarly distributed. 

The varied impact of the TV network on the cells is well adapted to by the \emph{proposed} approach. Compare the distributions (Figure~\ref{fig:CDFPower}) of power allocated to nodes in the cells by the \emph{proposed} and the \emph{baseline}. As shown earlier, nodes in cell $2$ have a large spread of link gains to the TV receiver. This leads to \emph{proposed} allocate a wide spread of transmit powers to them. The baseline, on the other hand, allocates all the nodes in the cell the same power. Nodes in cell $11$ too see a spread in power allocation on using \emph{proposed}. However, nodes in cell $36$ are assigned powers very similar to that assigned by baseline. Their being far from the TV network leads all nodes to see similar link gains to the TV receivers and also the transmitters. 
This ability of \emph{proposed}, which we illustrated using the cells $2$, $11$, and $36$, to adapt power allocation to link gains between nodes in a cell and between nodes and the receivers and transmitters of the TV network, explains the gains in throughput achieved by \emph{proposed} over \emph{baseline}.

\subsection{FCC Like Regulations} 
\label{sec:fcc_regulations}
Since \emph{proposed} and \emph{baseline} allocate transmit power to nodes in a manner such that interference constraints at TV receivers are not violated, nodes of a White-Fi cell can use a TV channel as long as they are outside the service contour of any TV transmitter broadcasting over the channel. FCC, instead, allows all nodes to use their full transmit power budget of $100$ mW as long as they are outside the protection contour. That is FCC regulations protect the TV receiver by simply ensuring that a separation distance is maintained between the afflicted receivers and the White-Fi nodes. While this method is simple as it doesn't require per node (\emph{Relaxed}) or per cell (\emph{Baseline}) power allocation, it reduces the region over which white space channels may be accessed by the White-Fi nodes. Also, a fixed separation distance can't ensure that the interference constraint at the TV receivers is satisfied for different White-Fi node densities.

Figure~\ref{fig:SepDis} shows the separation distance (from the service contour) that must be maintained for the constraints on maximum interference to be satisfied at the TV receivers while all nodes in the network use a fixed power of $100$ mW, as suggested by FCC regulations. Note that the protection contour provides for a separation distance of $11.1$ km~\cite{spectrumbridge}, which is not large enough for node densities larger than $1$ node/km$^2$ for Denver and for all chosen densities for the city of Columbus. Since aggregate interference is only a function of the number of nodes in the White-Fi network, the separation distance doesn't change with cell size. In summary, the separation distance, and hence the loss of coverage in white spaces, increases with increasing node density. Also, this distance is significantly large.

\subsection{Comments on Fairness and Overhead Rate}
\begin{figure*}[t]
\vspace{-8em}
  \centering
  \subfloat[]{\includegraphics[width=0.33\linewidth,bb = 0 0 950 900]{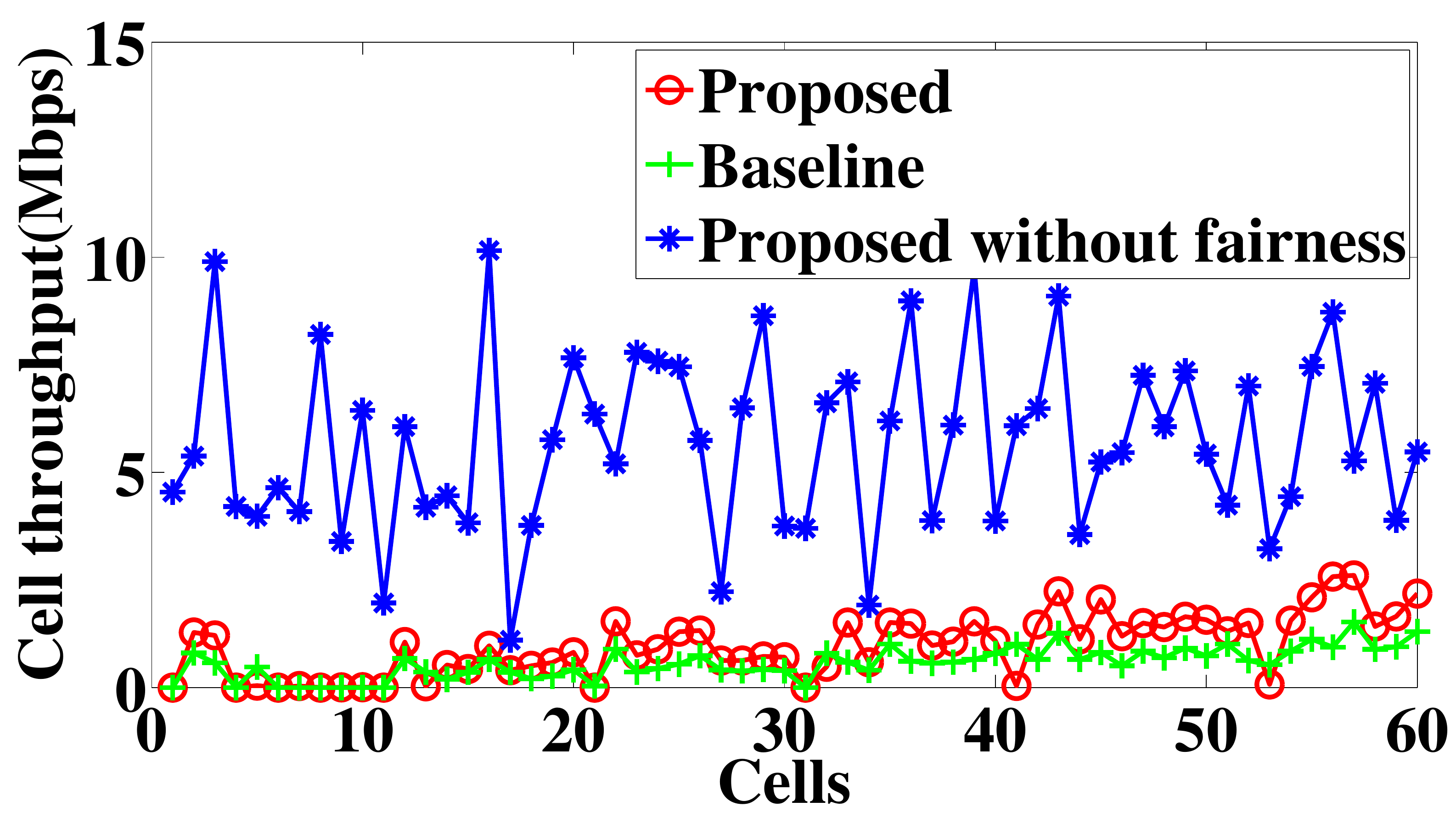}\label{fig:LinkThroughput}}
  \enspace
  \subfloat[]{\includegraphics[width=0.33\linewidth,bb = 0 0 950 900]{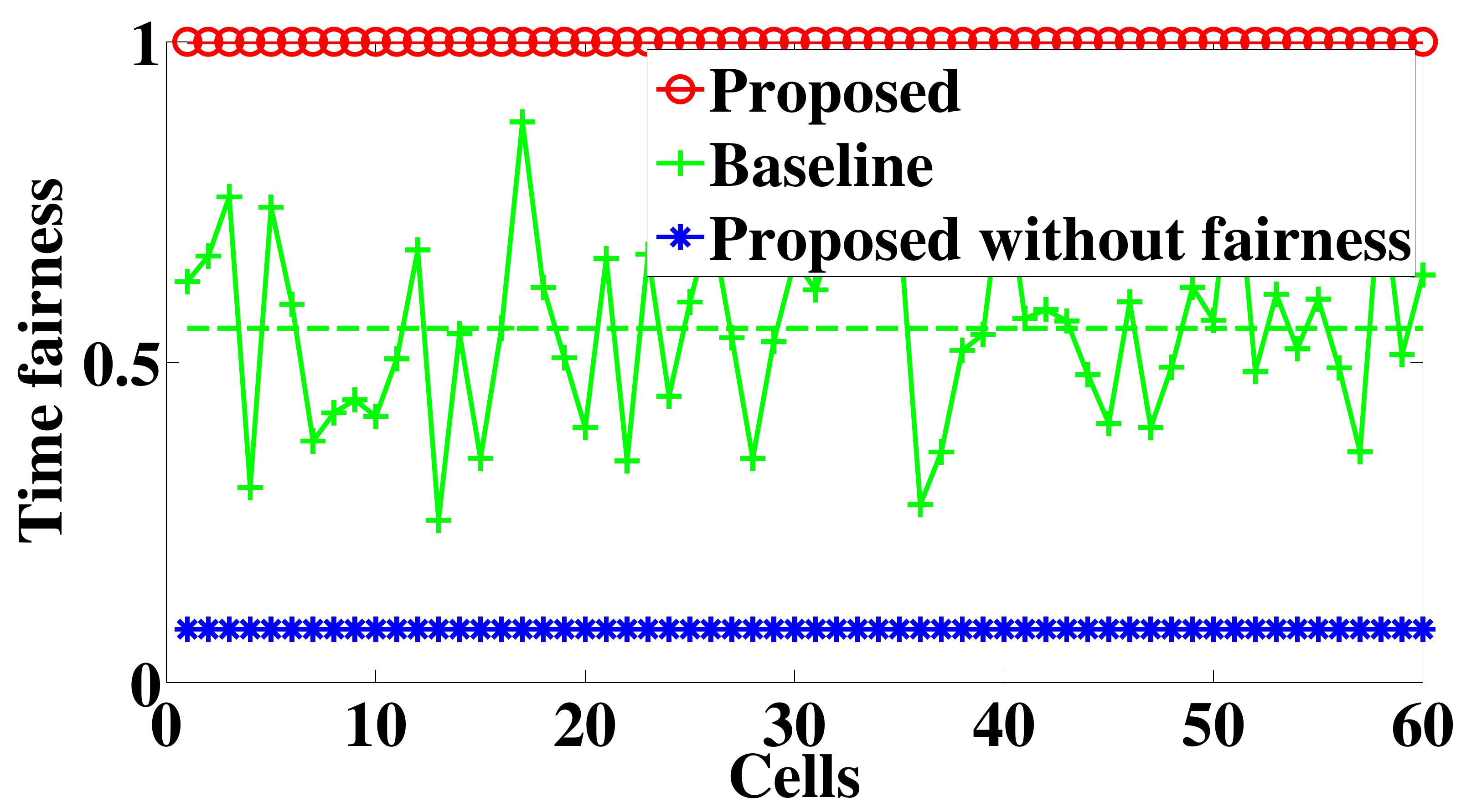}\label{fig:LinkTimeFairness}}
  \enspace
  \subfloat[]{\includegraphics[width=0.33\linewidth,bb = 0 0 950 900]{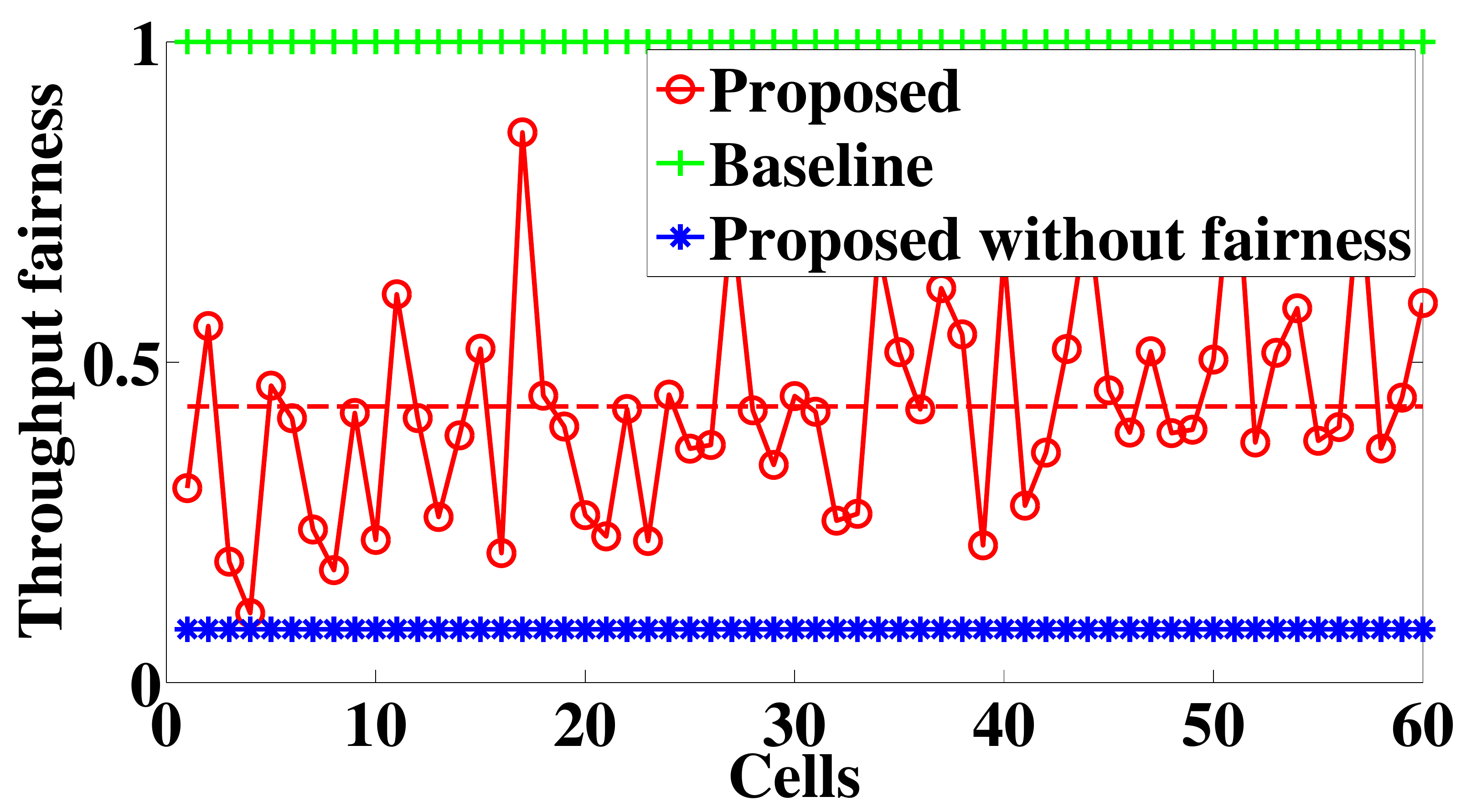}\label{fig:LinkThroughputFairness}}
  \caption{\small (a) Per cell throughput. (b) Per cell time fairness. (c) Per cell throughput fairness. Figure~\ref{fig:LinkThroughput}-\ref{fig:LinkTimeFairness} correspond to cells assigned channel $21$ in the White-Fi network in Denver with each cell covering an area of $12.25$ km$^{2}$. Results are shown for (i) \emph{Proposed} (ii) \emph{Baseline}, and (iii) \emph{Proposed} without fairness when using \textit{Relaxed}.}
  \label{fig:LinkFairness}
\vspace{-1em}
\end{figure*}

\begin{figure*}[t]
\vspace{-7em}
  \centering
  \subfloat[]{\includegraphics[width=0.25\linewidth, bb = 0 0 975 975]{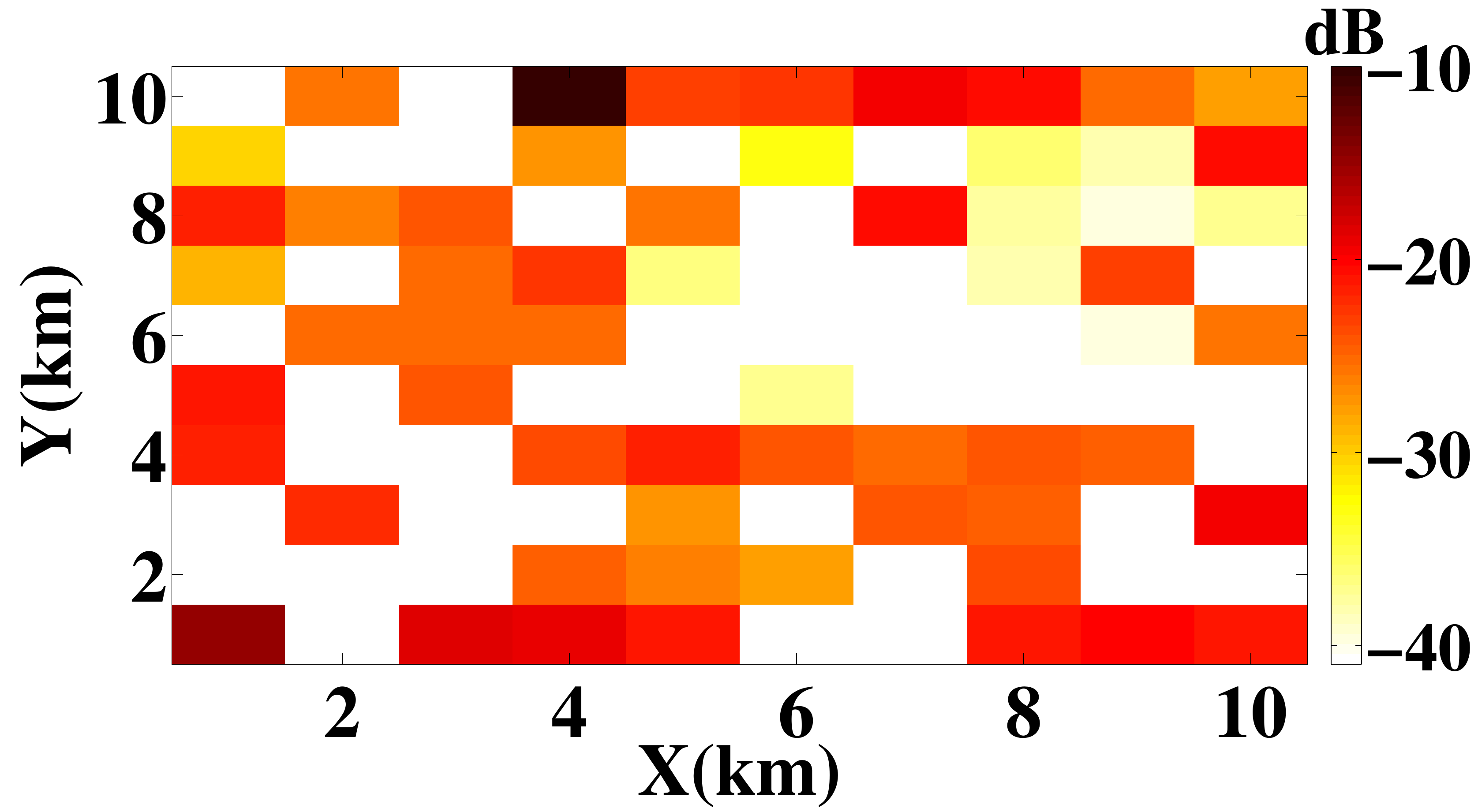}\label{fig:PowerAllocCell2}}
  \subfloat[]{\includegraphics[width=0.25\linewidth, bb = 0 0 975 975]{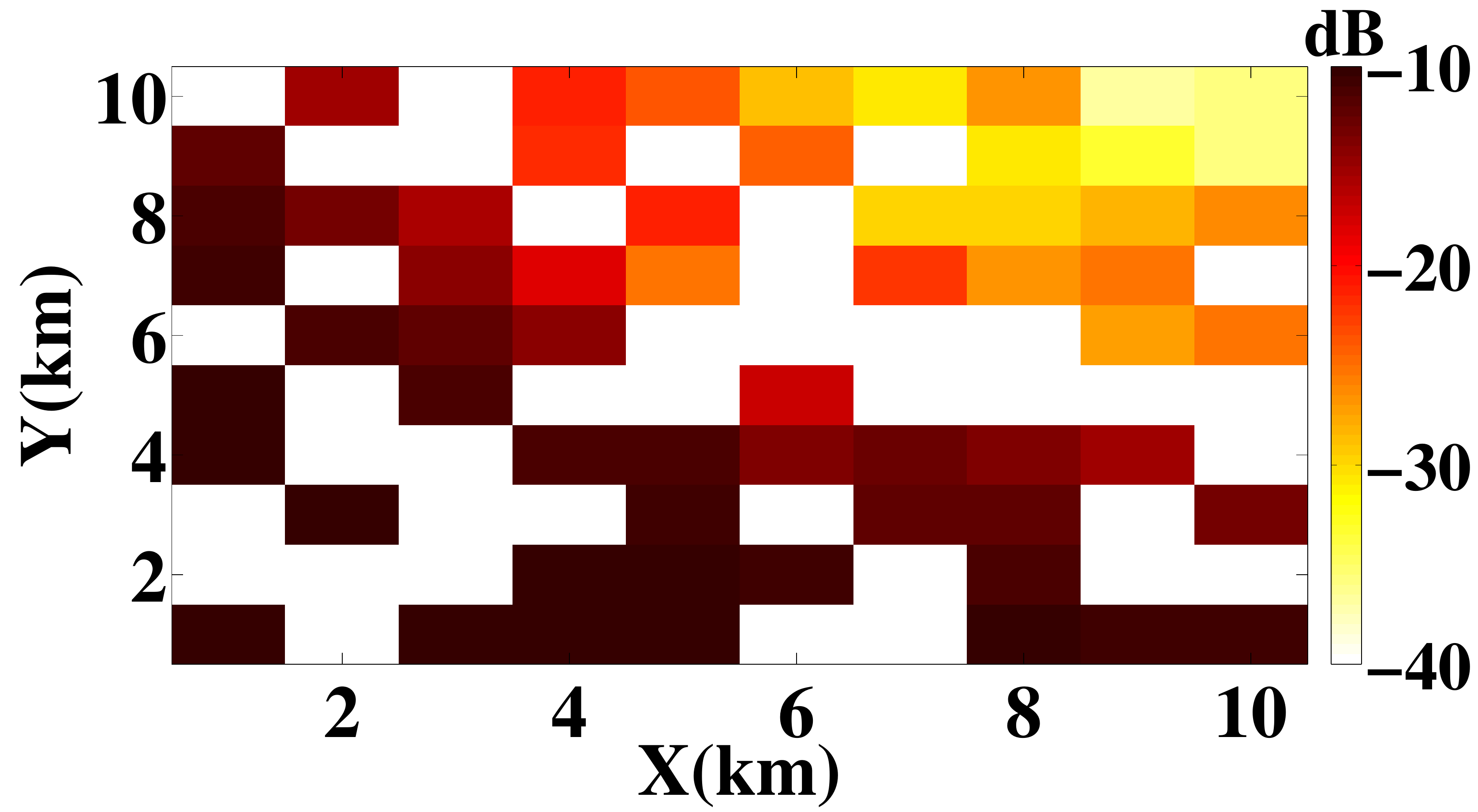}\label{fig:PowerAllocCell2NoOverheads}}
  \subfloat[]{\includegraphics[width=0.25\linewidth, bb = 0 0 975 975]{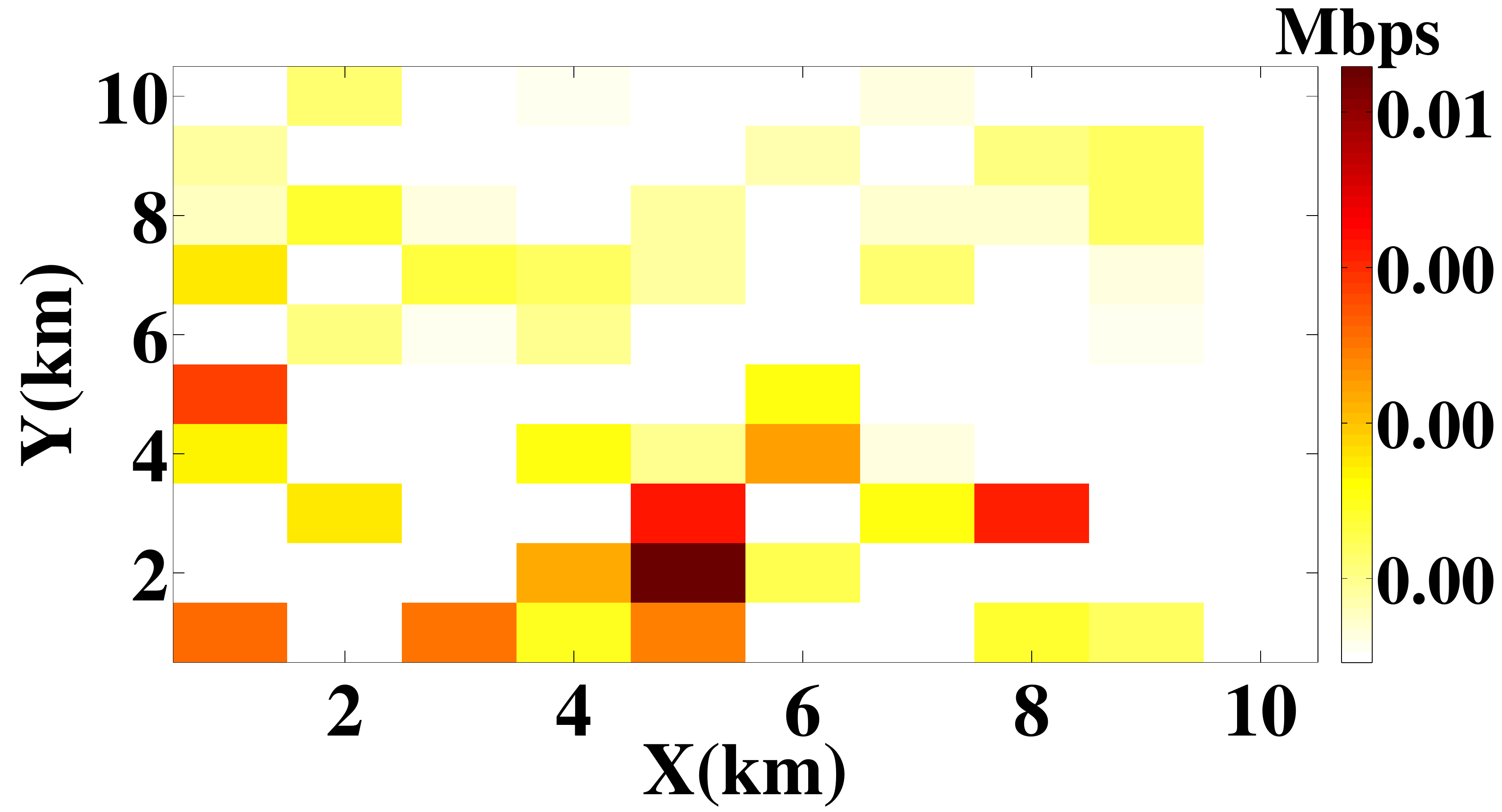}\label{fig:RateAllocCell2}}
  \subfloat[]{\includegraphics[width=0.25\linewidth, bb = 0 0 975 975]{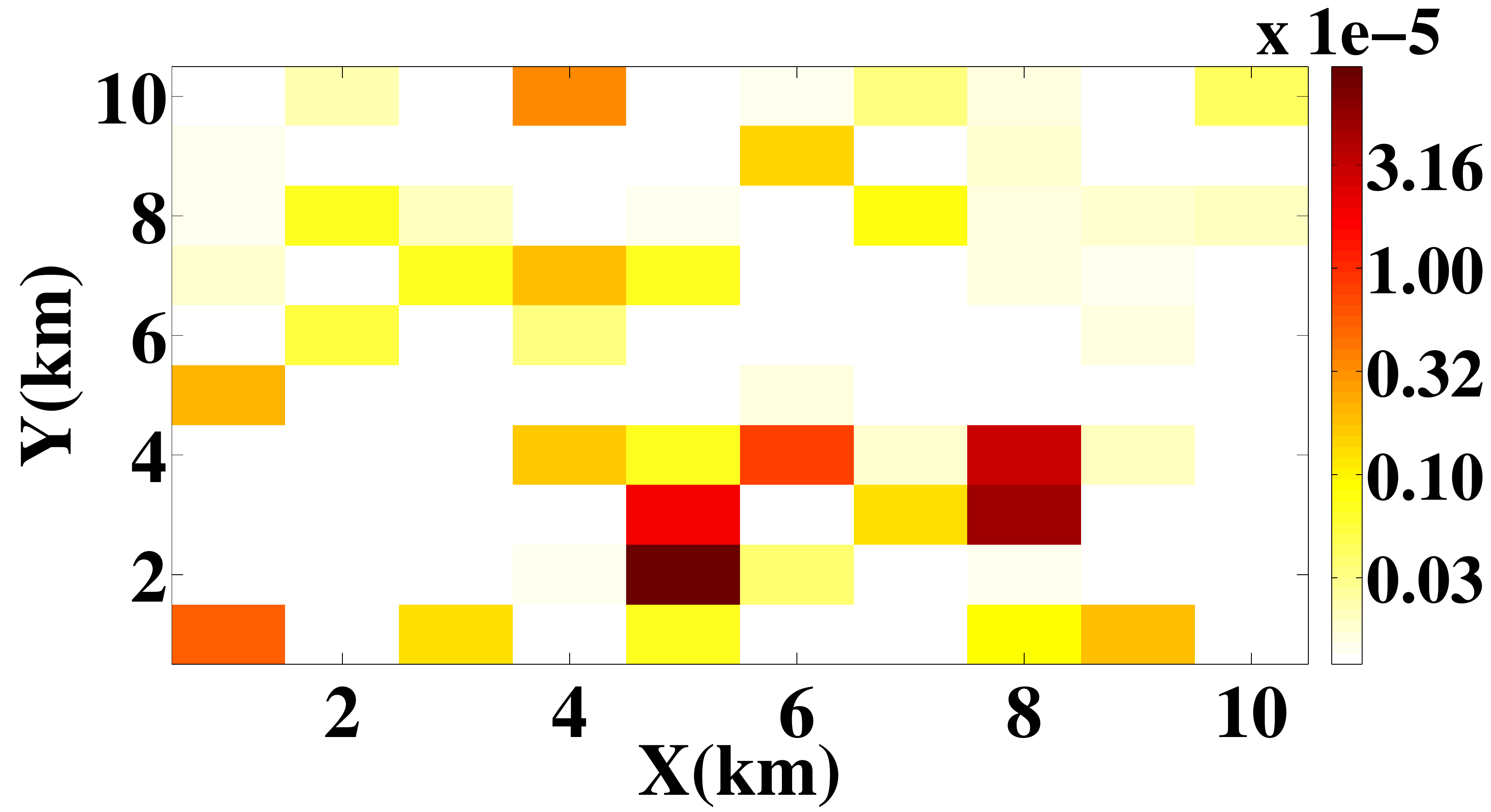}\label{fig:TauAllocCell2}}
  \caption{\small (a) Power allocated to White-Fi nodes, by the \emph{proposed} approach, in cell $2$ of the White-Fi network shown in Figure~\ref{fig:WorstReceivers}. (b) Power allocated as in (a), however, in the absence of overheads defined in Section~\ref{sec:model}. (c) Rates allocated to the nodes in cell $2$, corresponding to the power allocated in (a). (d) Access probabilities assigned to the nodes in cell $2$.}
\vspace{-1em}
\end{figure*}

\emph{Impact of the time fairness constraint:} We illustrate the impact of time fairness constraint (\ref{eqn:fairness}) in Figure~\ref{fig:LinkFairness} by comparing (a) \emph{proposed}, (b) \emph{baseline}, and (c) \emph{proposed} without the time fairness constraint being enforced. For each cell we define fairness in time share and fairness in throughput, obtained by links within the cell. We use Jain's fairness index~\cite{jain} to quantify fairness. The Jain's fairness index is defined as, 
$\mathcal{J} = \frac{\left(\sum_{i=1}^{n}x_{i}\right)^2}{n \sum_{i=1}^{n}x_{i}^2},$
where, $n$ is the number of links in a cell. For computation of fairness in time share, $x_{i}$ is the fraction of time $\frac{\pnodesuccPerChannel{s}{i} L/\RPerChannel{s}{i}{i'}}{\sigmaAvgPerChannel{s}}$ that the $i^{th}$ link spends on a successful transmission over channel $s$. 
For fairness in throughput, $x_i$ is the throughput $\frac{\pnodesuccPerChannel{s}{i} L}{\sigmaAvgPerChannel{s}}$ of the $i^{th}$ link.

As shown in Figure~\ref{fig:LinkThroughput}, the throughput obtained per cell for \emph{proposed} without time fairness is the largest. However, as shown in Figure~\ref{fig:LinkTimeFairness} and Figure~\ref{fig:LinkThroughputFairness}, this approach is highly unfair both in time share and throughput. This is because in the absence of time fairness only the link with the highest link quality in a cell accesses the medium while other links are starved. For the \emph{baseline}, as shown in Figure~\ref{fig:LinkThroughput}, it has the smallest per cell throughput and as shown in Figure~\ref{fig:LinkTimeFairness} an average time fairness of $0.55$ (obtained by  averaging the fairness indices across cells on the same channel). However, as shown in Figure~\ref{fig:LinkThroughputFairness}, it is highly throughput fair. The high throughput fairness in \emph{baseline} is because it assigns the same transmit power and access probabilities to all nodes in a cell. Lastly, as shown in Figure~\ref{fig:LinkTimeFairness}, the \textit{proposed} approach has high time fairness, but has an average throughput fairness of $0.43$ across cells on the same channel (see Figure~\ref{fig:LinkThroughputFairness}).

\emph{Rate allocation and access probability assignment under fairness constraint:}
Figures~\ref{fig:RateAllocCell2} and~\ref{fig:TauAllocCell2} respectively show the rate and access probability assignment for White-Fi nodes in cell $2$. Nodes assigned larger payload rates (because of larger SINR) have larger access probabilities. This is because we optimize under the constraint~(\ref{eqn:fairness}) of time fairness. For the case when all nodes have White-Fi links with the same SINR to their destinations, we confirm that all nodes transmit at the same rates and use the same access probability per channel. In fact, the access probabilities are the same as those shown via simulation and approximate analysis in~\cite{bianchi}.

\emph{Impact of Overhead Rate On Power Allocation:} Recall from Section~\ref{sec:model} that White-Fi nodes transmit overheads at a rate that all nodes in their cell can decode. Figure~\ref{fig:PowerAllocCell2} shows the power allocation to nodes in the White-Fi cell marked $2$ in Figure~\ref{fig:WorstReceivers}. Each node in the cell is colored in accordance with power allocated to it. While the high power allocation to nodes that are far from the TV receiver (nodes closer to $(0,0)$ in Figure~\ref{fig:PowerAllocCell2}) is as per expectation given that such nodes have smaller link gains to the TV receiver and transmitter, the high power allocation to nodes that are closest to the TV receiver (nodes closer to $(10,10)$) is explained by the requirement of the overhead rate.

This rate is determined by the nodes in the White-Fi cell that are farthest from each other. Allocating a small power to nodes that are close to TV receiver will make the rate very small, which in turn will adversely affect the throughputs of all nodes in the cell. Power allocation of the kind seen in Figure~\ref{fig:PowerAllocCell2} is seen in cells in which nodes are unable to use their entire power budget because of the constraint~(\ref{eqn:maxConst}) on aggregate interference.

Figure~\ref{fig:PowerAllocCell2NoOverheads} shows the power allocation for the cell $2$ when the number of overhead bits is forced to zero. The lack of overhead bits makes the overhead rate inconsequential. The resulting power allocation follows the familiar pattern~\cite{sahai} of the allocation being larger at nodes that are farther from the TV receiver.

\section{Conclusions}
\label{sec:conclusions}
\SG{We modeled the saturation throughput of a city-wide White-Fi network. We demonstrated heterogeneity in channel availability and link quality in such networks and proposed a method to optimize the saturation throughput that effectively leveraged the heterogeneities. We demonstrated the efficacy of our method using hypothetical deployments of White-Fi networks amongst real TV networks in the cities of Denver and Columbus. This led to the observation that high channel availability may not translate to high network throughput. We compared the proposed approach with a simpler to implement baseline and FCC-like mechanisms. Our approach showed significant gains over the baseline ($40-70\%$). We also evaluated, both the proposed approach and the baseline, for scenarios when white space channels can be accessed by nodes anywhere outside (a) the protected region (as per FCC regulations), referred as \textit{Exact} FCC, and (b) the service contour, referred as \textit{Relaxed}. The choice of \textit{Relaxed} showed a gain of $27\%$ over \textit{Exact} FCC in Denver and $36\%$ in Columbus for the proposed approach, hence favoring the elimination of protected regions.}

\section{Acknowledgements}
\label{sec:ack}
This work has been partly supported by DeitY, Government of India, under grant with Ref. No. R-23011/1/2014-BTD. We thank the reviewers and Dr. Pravesh Biyani, IIIT-Delhi, for their valuable comments.
\begin{spacing}{0.9}
\bibliographystyle{IEEEtran}
\bibliography{references}
\end{spacing}
\appendices
\section{Solve for Access}
\label{appendix:A}
We show how the problem of maximizing $\throughputCellCh{m}{s}$ can be reduced to a minimization problem~(\ref{eqn:access1})-(\ref{eqn:access2}) in a single-variable $\accessPerChannel{s}{j}$ of node $j$, where $j$ is the node in cell $m$ whose payload rate $\RPerChannel{s}{j}{j'}$ on channel $s$ is the smallest amongst all nodes in the cell. We also show that the resulting problem~(\ref{eqn:access1})-(\ref{eqn:access2}) is convex in $\accessPerChannel{s}{j}$.

The throughput of the $i^{th}$ node in cell $m$ on a channel $s$ assigned to it is given as
$\throughputCellCh{i,m}{s} =  \frac{\pnodesuccPerChannel{s}{i} L}{\sigmaAvgPerChannel{s}}.
\label{eqn:perCellPerChannelPerNodeTh}$
On simplifying $\throughputCellCh{i,m}{s}$ we get
{\small{
\begin{align}
\throughputCellCh{i,m}{s} &= \left[\frac{1}{L}\left(\frac{\pidle}{\pnodesuccPerChannel{s}{i}}\sigma +\Tnodesucc{i} + \left(\sum\limits_{\substack{j\in \mathcal{N}_m \\ j\neq i}}\frac{\pnodesuccPerChannel{s}{j}}{\pnodesuccPerChannel{s}{i}}\Tnodesucc{j}\right)\right.\right.\nonumber\\
&\enspace\left.\left.+\left(\frac{1}{\pnodesuccPerChannel{s}{i}}-\frac{\pidle}{\pnodesuccPerChannel{s}{i}}-1-\sum\limits_{\substack{j\in \mathcal{N}_m \\ j\neq i}}\frac{\pnodesuccPerChannel{s}{j}}{\pnodesuccPerChannel{s}{i}}\right)\Tcol\right)\right]^{-1},
\end{align}
}}
where the probability $\pnodesuccPerChannel{s}{i}$ that a transmission by $i$ is successful and the probability $\pidle$ that a slot is idle is given in~(\ref{eqn:psuccAndIdleForANode}). Using the time fairness constraint~(\ref{eqn:fairness}), we can write $\pnodesuccPerChannel{s}{i}\RPerChannel{s}{j}{j'} = \pnodesuccPerChannel{s}{j}\RPerChannel{s}{i}{i'}$, which gives
\begin{align}
\accessPerChannel{s}{i} = \frac{\accessPerChannel{s}{j}\RPerChannel{s}{i}{i'}}{(1-\accessPerChannel{s}{j})\RPerChannel{s}{j}{j'} + \accessPerChannel{s}{j}\RPerChannel{s}{i}{i'}}.\label{eqn:access}
\end{align}
By substituting $\pnodesuccPerChannel{s}{i}$ and $\pidle$ from~(\ref{eqn:psuccAndIdleForANode}) and $\accessPerChannel{s}{j}$ from~(\ref{eqn:access}) we get
{\footnotesize{
\begin{align}
\throughputCellCh{i,m}{s} &= \left[f_{1}(P) + \frac{\RPerChannel{s}{j}{j'}}{L\RPerChannel{s}{i}{i'}} \left(\frac{1-\accessPerChannel{s}{j}}{\accessPerChannel{s}{j}}\sigma+\left(\prod\limits_{k\in\mathcal{N}_m}\left(\frac{\RPerChannel{s}{k}{k'}}{\RPerChannel{s}{j}{j'}} +\frac{1-\accessPerChannel{s}{j}}{\accessPerChannel{s}{j}}\right)\right.\right.\right.\nonumber\\
&\hspace{11em}\left.\left.\left.-\frac{1-\accessPerChannel{s}{j}}{\accessPerChannel{s}{j}}-f_{2}(P)\right)\Tcol\right)\right]^{-1},
\end{align}
}}
where, $f_{1}(P)= \frac{\Tnodesucc{i}}{L} + \left(\sum\limits_{\substack{j\in \mathcal{N}_m \\ j\neq i}}\frac{\RPerChannel{s}{j}{j'}}{L\RPerChannel{s}{i}{i'}}\Tnodesucc{j}\right),\enspace f_{2}(P)= \frac{\RPerChannel{s}{i}{i'}}{\RPerChannel{s}{j}{j'}}\left(1+\sum\limits_{\substack{j\in \mathcal{N}_m \\ j\neq i}}\frac{\RPerChannel{s}{j}{j'}}{\RPerChannel{s}{i}{i'}}\right).$
Observe that $f_{1}(P)$ and $f_{2}(P)$ are independent of $\accessPerChannel{s}{j}$ and do not affect the optimal $\accessPerChannel{s}{j}$. This allows us to reduce the throughput maximization problem to the problem~(\ref{eqn:access1})-(\ref{eqn:access2}).

Next we show that~(\ref{eqn:access1})-(\ref{eqn:access2}) is a convex optimization problem. 
Define $f(\accessPerChannel{s}{j})$ to be the utility function (Equation~(\ref{eqn:access1})) in the problem~(\ref{eqn:access1})-(\ref{eqn:access2}). The function $f(\accessPerChannel{s}{j})$ may be written as
\begin{align}
& f(\accessPerChannel{s}{j}) = \sigma f_{1}(\accessPerChannel{s}{j})+\Tcol f_2(\accessPerChannel{s}{j}),
\label{eqn:funcTau}
\end{align}
where $f_{1}(\accessPerChannel{s}{j}) = \frac{1-\accessPerChannel{s}{j}}{\accessPerChannel{s}{j}},\ f_{2}(\accessPerChannel{s}{j}) = \prod\limits_{k\in \mathcal{N}_m }\left(\frac{\RPerChannel{s}{k}{k'}}{\RPerChannel{s}{j}{j'}} +\frac{1-\accessPerChannel{s}{j}}{\accessPerChannel{s}{j}}\right)-\frac{1-\accessPerChannel{s}{j}}{\accessPerChannel{s}{j}}.$
$f_{1}(\accessPerChannel{s}{j})$ is a non-increasing convex function on the interval $[0,1]$. Next, we show that the function $f_{2}(\accessPerChannel{s}{j})$ is also a convex function. For this, we rewrite $f_{2}(\accessPerChannel{s}{j})$ as
\begin{align}
f_{2}(\accessPerChannel{s}{j})
& = \frac{1}{\accessPerChannel{s}{j}}\left(\prod\limits_{\substack{k\in \mathcal{N}_m \\ k\neq j}} \left(\frac{\RPerChannel{s}{k}{k'}}{\RPerChannel{s}{j}{j'}} +\frac{1-\accessPerChannel{s}{j}}{\accessPerChannel{s}{j}}\right)-1\right)+1,
\end{align}
and use the following fact~\cite[Chapter~3, Question~3.32]{boyd}.
\begin{lemma}
If $f:\mathcal{R}\rightarrow \mathcal{R}:x\mapsto f(x)$ and $g:\mathcal{R}\rightarrow \mathcal{R}:x\mapsto g(x)$ are both convex, non-decreasing (or non-increasing) and positive, then $h:\mathcal{R}\rightarrow \mathcal{R}:x\mapsto h(x) = f(x)g(x)$ is also convex.\label{lem:Lemma1}
\end{lemma}
Observe that $f_{2}(\accessPerChannel{s}{j})$ consists of a product of $\frac{1}{\accessPerChannel{s}{j}}$ and $\left(\prod\limits_{\substack{k\in \mathcal{N}_m \\ k\neq j}} \left(\frac{\RPerChannel{s}{k}{k'}}{\RPerChannel{s}{j}{j'}} +\frac{1-\accessPerChannel{s}{j}}{\accessPerChannel{s}{j}}\right)-1\right)$. Both the functions are convex, non-increasing and positive functions on the interval $[0,1]$. Hence, using Lemma~\ref{lem:Lemma1}, the product of these functions i.e. $f_{2}(\accessPerChannel{s}{j})$ is also convex on the interval $[0,1]$.

Therefore, $f(\accessPerChannel{s}{j})$ in~(\ref{eqn:funcTau}) is a non-negative weighted sum of convex functions i.e. $f_{1}(\accessPerChannel{s}{j})$ and $f_{2}(\accessPerChannel{s}{j})$, and hence a convex function~\cite{boyd}. Also note that the constraint set given by~(\ref{eqn:access2}) is convex. Thus the problem~(\ref{eqn:access1})-(\ref{eqn:access2}) is a convex optimization problem.
\end{document}